\documentclass[12pt,preprint]{aastex}

\usepackage{CJK}
\usepackage{graphics,epsfig}
\usepackage{natbib}
\usepackage{lscape}

\def\spose#1{\hbox to 0pt{#1\hss}}
\def\lta{\mathrel{\spose{\lower 3pt\hbox{$\mathchar"218$}}
     \raise 2.0pt\hbox{$\mathchar"13C$}}}
\def\gta{\mathrel{\spose{\lower 3pt\hbox{$\mathchar"218$}}
     \raise 2.0pt\hbox{$\mathchar"13E$}}}
\def\eq1{$\equiv$~1}

\def\figure#1#2 {\par{\narrower\noindent {\bf Fig. #1}
   \hskip 2mm #2\par}\bigskip\noindent}
\def\table#1#2 {\par{\narrower\noindent {\bf Tab. #1}
   \hskip 2mm #2\par}\bigskip\noindent}

\def\registered{{\ooalign{\hfil\raise .00ex\hbox{\scriptsize R}\hfil\crcr\mathhexbox20D}}}

\usepackage{amsmath}
\usepackage{accents}
\newlength{\dhatheight}

\slugcomment{Version: \enskip \today}

\shorttitle{Habitability in Binary Systems III}

\shortauthors{Cuntz et al.}

\begin{document}
\begin{CJK*}{UTF8}{gbsn}


\title{{\it S}-Type and {\it P}-Type Habitability in Stellar Binary Systems: \\
A Comprehensive Approach \\
III. Results for Mars, Earth, and super-Earth Planets
}

\author{Zh. Wang (王兆鹏) and M. Cuntz}

\affil{Department of Physics}
\affil{University of Texas at Arlington, Arlington, TX 76019-0059;}
\email{zhaopeng.wang@mavs.uta.edu; cuntz@uta.edu}

\begin{abstract}
In Paper~I and II, a comprehensive approach was utilized for the calculation of 
{\it S}-type and {\it P}-type habitable regions in stellar binary systems
for both circular and elliptical orbits of the binary components.  It
considered a joint constraint including orbital stability and a habitable
region for a possible system planet through the stellar radiative energy
fluxes (``radiative habitable zone"; RHZ).  Specifically, the stellar
{\it S}-type and {\it P}-type RHZs are calculated based on the solution
of a fourth order polynomial.  However, in concurrent developments, mostly
during 2013 and 2014, important improvements have been made in the computation
of stellar habitable zones for single stars based on updated climate models
given by R. K. Kopparapu and collaborators.  These models entail considerable
changes for the inner and outer limits of the stellar habitable zones.
Moreover, regarding the habitability limit given by the runaway greenhouse
effect, notable disparities were identified between Earth, Mars, and
super-Earth planets due to differences in their atmospheric models, thus
affecting their potential for habitability.  It is the aim of this study
to compute {\it S}-type and {\it P}-type habitable regions of binaries in
response to the updated planetary models.  Moreover, our study will also
consider improved relationships between effective temperatures, radii,
and masses for low-luminosity stars.
\end{abstract}



\section{INTRODUCTION}

Do habitable zones\footnote{A glossary of acronyms is given in Appendix~A.}
(HZs) exist only around single stars?  The answer is certainly ``no",
according to many years of research pursued by different groups.  Previous theoretical results
about zones of possible habitability for binary (and higher-order) systems have been obtained
by, e.g., \cite{egg12,egg13}, \cite{kan13}, \cite{kal13}, \cite{hag13}, \cite{cun14,cun15}, and \cite{baz17}, among
others.  The main goal of these studies was the identification of stellar binary systems with
respect to different types of planetary and stellar configurations.  For example, \cite{baz17}
explored the dynamics and habitability in circumstellar planetary systems of previously
identified binaries including the consideration of resonance phenomena.

The overall importance of studying stellar binary systems stems from their relatively high
frequency \citep[e.g.,][]{duq91,pat02,egg04,lad06,rag06,rag10,roe12}.  In fact, following
those works, the majority of stars, notwithstanding M-type dwarfs, are not single.
Additionally, a significant number of binary systems have also been identified to harbor
planets.  Following \cite{dvo82}, there are two kinds of systems.  First, planets may orbit
one of the binary components; those are said to be in an {\it S}-type orbit.  In those systems
the other stellar component is at a notable distance; however, it may act as a perturbator
--- possibly resulting in considerable orbital ramifications.  Second, planets may orbit
both binary components; those are said to be in {\it P}-type orbits.  In summary, so far close
to 100 planet-hosting binary systems have been identified.  A survey on planet-hosting
exoplanetary systems, with focus on stellar separations of less than 100~au, has been
published by \cite{baz17}.  Studies about the successful formation of planets in binary systems
have been given by, e.g., \cite{kle12}, \cite{the15}, and references therein.

In Paper~I and II of this series, \cite{cun14,cun15} focused on a theoretical identification
(i.e., ab-initio approach) of {\it S}-type and {\it P}-type HZs in binary systems, including the
possibility of no HZ after all.  The approach-as-elected included (1) the consideration of a
joint constraint including orbital stability and a habitable region for a possible system planet
through the stellar radiative energy fluxes needs to be met; (2) the treatment of different
types of HZs; and (3) the provision of a combined formalism for the assessment of both
{\it S}-type and {\it P}-type HZs based on the solution of a quartic algebraic equation.
\cite{cun15} presented an expansion of the methodology to binary systems based on
elliptical orbits.

A critical aspect pertaining to the calculation of binary HZs is the choice of planetary
climate models.  As part of a quasi-contemporaneous development, \cite{kop13,kop14} offered
new estimates for the widths and extents of HZs around single stars based on significant
improvements of the previous work by \cite{kas93}.  They constructed an updated
1-D radiative--convective, cloud-free climate model based on new data (including,
but not limited to, improved H$_2$O and CO$_2$ cross sections), which allowed them to
calculate revised stellar HZs for F, G, K, and M stars.  For one of the climate models,
\cite{kop14} identified a weak, but nonetheless identifiable, dependency on the planetary
mass.  In that case, they constructed models appropriate for Mars-type, Earth-type
(the previous default), and super-Earth-type planets.  This discrepancy has been obtained
for the entire set of stars.

In this study we discuss updated results about the structure and existence of HZs in stellar binaries.
Our paper is structured as follows.  In Section~2, we describe our theoretical approach with focus on
the key equations.  We also comment on the set of stellar parameters used and on the revised habitable zone
estimates.  In Section~3, we describe our results and discussion.  Our summary and conclusions are given
in Section~4.


\section{THEORETICAL APPROACH}


\subsection{Key Equations}



The key equations of this study closely follow the previous work given as Paper~I and II, and references
therein.  The habitability limits around single stars can be calculated using
\begin{equation}
d^2 \ = \ \frac{L/L_\odot}{S_{{\rm eff}, {\ell}}}
\end{equation}
with $L/L_{\odot}$ as the star's luminosity in units of the solar luminosity, and $S_{{\rm eff}, \ell}$ as
effective stellar flux, in unit of solar constant.  The latter is a function of stellar effective temperature,
i.e., $S_{{\rm eff}, \ell} = S_{{\rm eff}, \ell}(T_{\rm eff})$ that also depends on the type of HZ limit
$\ell$ (see Section~2.2 for details and discussion.  For the Sun itself, the following relationship holds
\begin{equation}
s_{\ell} \ = \ {S_{{\rm eff}, {\ell}}}^{-{1/2}}
\end{equation}
with $s_{\ell}$ denoting the various inner/outer limits of the solar HZ.

As focus of our study, we adopt the habitability limits based on the climate models by
\cite{kop14}.  They found that the runaway greenhouse effect, readily taken as the inner limit
for the general habitable zone (GHZ), also depends on the planetary mass.  Thus, we
introduce {\it k} stands for the type of planet (see Section~2.2).  Therefore, we adopt a
more generalized notation given as $s_{{\ell}} \rightarrow s_{{\ell}k}$ and
$S_{{\rm eff}, {\ell}} \rightarrow S_{{\rm eff}, {\ell}k}$.

Generally, the distinctions between habitability limits of different stars are conveyed through the
different choices of stellar luminosity $L$, effective temperature $T_{\rm eff}$, type of habitability limit $\ell$
and planetary mass, if applicable (parameterized through $k$).  In case of (theoretical) main-sequence stars,
the number of parameters is further reduced due to the relation between $L$ and $T$.  Interpolations
for $S_{{\rm eff}, {\ell}k}$ are available; see Appendix~B and Figure~2 for details.  In the following,
we use the approach by \cite{kop13,kop14}, except if noted otherwise.

For multiple stellar systems with $N$ stars, by taking the radiation from all stellar components into account,
the radiative habitable zone (RHZ) --- which purposely ignores any orbital stability requirement for
possible system planets --- 
can be calculated following the inverse-square law.  Hence, the key equation reads
\begin{equation}
\sum_{i=1}^N \frac{L_i/L_\odot}{{S_{{\rm eff}, i{\ell}k}} \ d_i^2} \ = \ 1
\end{equation}
Akin to Paper~I, we define $L_{i{\ell}k}'$ as
\begin{equation}
L_{i{\ell}k}' = \frac{L_i}{L_{\odot} {S_{{\rm eff}, i{\ell}k}}}
\end{equation}
Consequently, the key equation for a binary star system reads
\begin{equation}
\sum_{i=1}^2 \frac{L_{i{\ell}k}'}{d_i^2} \ = \ 1 \ ;
\end{equation}
see Figure~1 for information on the system set-up.

The RHZ of the binary system could be found through solving some lengthy algebra (see Paper~I and II)
involving the solution of a quartic equation\footnote{In Eq.~(20) of Paper~I, the third term should correctly
read $-4 A_0 y$.  However, this typo does not affect any of the calculations in Paper~I or subsequent work.
}; see Figure~1 for information on the mathematical set-up.  Furthermore, the limit of orbital
stability needs to be taken into account as well in order to obtain viable stellar HZs,
which is done following the work\footnote{
Updated results for planetary orbital stability concerning both {\it S-}type and {\it P-}type
binary systems have been given by \cite{qua18}.  However, in those simulations a Jupiter-mass
planet has been used as test object to determine the planetary orbital stability limits.
Hence, tighter limits have been identified compared to the work by \cite{hol99} also used here.
However, in the view of the uncertainty bars of the latter work both kinds of results are
mutually consistent.
}
by \cite{hol99}. The orbital stability limit $a_{\rm cr}$, which depends on the binary separation
$a_{\rm bin}$, binary eccentricity $e_{\rm bin}$ and stellar masses $M_1$ and $M_2$, is an {\it upper limit}
for {\it S}-type system centered from the primary star,  and a {\it lower limit} for {\it P}-type system
measured from the binary mass center.  Figure~3 depicts a flow diagram on the newly devised online tool
{\tt BinHab 2.0} allowing for the computation of the existence of HZs in binary systems.  Following the
nomenclature of Paper~I, if {\it S-}type and {\it P-}type HZs are {\bf T}runcated because of the
criterion of orbital stability for system planets, the terms {\it ST-}type and {\it PT-}type HZs
are used.


\subsection{Stellar Parameters and Habitable Zone Estimates}



For this study, stellar parameters are required for which we adopt standard
main-sequence, see, e.g., \cite{bar98} and \cite{gra05}.  A notable exception,
however, are stars of spectral type K and M, for which we consider updated
research results; see Table~1 for data on stellar parameters\footnote{In
Paper~I and II, the Sun was equated to a theoretical main-sequence stars
of $T_{\rm eff} = 5811$~K following \cite{gra05}.  In the present work,
we assume $T_{\rm eff} = 5780$~K,  a value (within its uncertainty bar)
identical to the observed value of 5777~K \citep{sti04}, as also used in
the climatological studies by \cite{kop13,kop14}.} for stars
in the spectral range between F0 and M2.  Here we also utilize results by
\cite{man13}, including their mass -- effective temperature and mass --
radius relationships as implied.  Table~1 conveys the stellar parameters
as adopted here ranging from spectral type F0 to M2.

\cite{man13} analyzed moderate resolution spectra for a set of nearby K and M dwarfs
with well-known parallaxes and interferometrically determined radii to define
their effective temperatures, among other quantities.  As part of their
efforts, they focused on the hosts of transiting planet candidates in the
{\it Kepler} field to determine the stellar masses and to place additional
constraints to the remaining stellar parameters.
Generally speaking, for standard (theoretical) main-sequence stars as considered here,
the stellar luminosities and masses are closely correlated, which reduces the number
of free stellar and system parameters from six (i.e., $M_1$, $M_2$,
$L_1$, $L_2$, $a_{\rm bin}$, $e_{\rm bin}$) to four as the mass and luminosity
shall be viewed as mutually redundant.  The reason is that for those stars the
latter are connected by mass-luminosity relationships; see, e.g., \cite{rei87},
and references therein.  An updated theoretical mass-luminosity relation for K
and early M dwarfs has recently been given by \cite{cun18}.

Another important aspect of our study consists in obtaining adequate HZ estimates.
An early version of this concept has been introduced by \cite{kas93} [Kas93].
As inner limits of HZs, they used criteria following conditions akin to Recent
Venus (RV), the runaway greenhouse effect, and the moist greenhouse effect.
As outer limits of HZs, they used criteria following conditions akin to Early
Mars (EM), the maximum greenhouse effect (no clouds), and the first CO$_2$ condensation;
Furthermore, following \cite{kas93}, the occurrence of the
runaway greenhouse effect means that the greenhouse phenomenon is enhanced
by water vapor, resulting in surface warming.  As indicated by the models,
that effect further increases the atmospheric vapor content, leading to an
additional rise of the planetary surface temperature.

Furthermore, water loss criterion means that an atmosphere is warm enough to
have a wet stratosphere, from where water is gradually lost by atmospheric
chemical processes to space.  The first CO$_2$ condensation, a criterion
later abandoned by \cite{kop13,kop14} [Kop1314], indicates the stellar distance
where CO$_2$ start to form, thus significantly shaping the planetary
climate conditions at larger distances as well; see Table~2 for comparisons
for habitability limits pertaining to the Sun, $s_{\ell}$, as used by different
authors.  Regarding the runaway greenhouse limit, \cite{kop14} also found
a weak, but nonetheless identifiable, dependency on the planetary mass.
Their models allowed to calculate limits for Mars-type ($M = 0.1~M_\oplus$),
Earth-type ($M = M_\oplus$), and super-Earth-type ($M = 5.0~M_\oplus$) planets
considered here as well.  Regarding index $\ell = 2$, denoting the
limit due to the runaway greenhouse effect (an inner limit), an additional
index $k$ is used, with $k=1,0,2$ indicating an Earth-type, Mars-type, or
super-Earth-type planet, respectively.

These definitions allow to define the limits for the RVEM, GHZ,
as well as conservative habitable zone (CHZ); see Table~3 for details
including comparisons with previous work.  In Paper I and II, we largely followed
the conventions proposed by \cite{kas93}, except we did not consider
the RVEM HZ.  Instead, we utilized the extended habitable zone (EHZ)\footnote{
The work by \cite{mis00}, not used here, but considered in Paper~I and II, has
been superseded by more recent studies, including work by \cite{hal09}, \cite{pie11},
\cite{kit16}, and \cite{ram17,ram18}.  \cite{kit16} argued that the heating
assumed by \cite{mis00} has been overestimated, thus putting the extension
of the outer HZ in question.  \cite{pie11} pointed out that planetary HZs could
actually extend to up to 10~AU for solar-like stars.  Additional results,
including those for methane-based HZs, have been given by \cite{ram17,ram18}.}
introduced by \cite{mis00}.  The outer limit of the EHZ signifies
the maximum greenhouse effect, assuming 100\% cloud coverage, based on a simplified
model available at the time.  It is noteworthy that there are no universally accepted
definitions what constitutes the CHZ and GHZ, as those tend to vary from author-to-author.
In Table~3, we list the ones adopted by \cite{kop13,kop14}, as well as those used in
the present work where we use the conventions GHZ and RVEM.  Furthermore, Table~4
conveys the main target list for this study, encompassing stars of masses
1.25, 1.00, 0.75, and 0.50~$M_\odot$, corresponding to (approximate)
spectral types of F8~V, G2~V, K2~V, and M1~V, respectively.  Stellar data, i.e.,
effective temperature $T_{\rm eff}$ and radius $R_\ast$, are obtained through
interpolation for set values of stellar mass.
Regarding stellar luminosity, this selection of stars corresponds
to a range between 2.15~$L_\odot$ and $3.59 \times 10^{-2}$~$L_\odot$.

The acquisition of updated stellar parameters, based either on theory or observations,
is an ongoing process.  Especially the derivation of mass-luminosity-radius-spectral type
relationhips for M-dwarfs has proven to be particularly challenging; for information on
current research see, e.g., \cite{vey17} and references therein.  Although most of
our study utilizes the stellar parameters of \cite{man13}, as an example of comparison
with other data, we calculate selected models for both {\it P-} and {\it S-}type
habitability for middle M dwarfs based on the data of \cite{bar15}.  These authors
have provided evolutionary models for pre-main-sequence and main-sequence
low-mass stars while utilizing updated data bases.  The differences for the HZs
as obtained are relatively minor, and are most pronounced for pairs of M-dwarfs; see
Section 3.2.3 for details.


\subsection{Comments on the Coordinate System}



Regarding the coordinate system (COS) used for our study, the semi-distance between the
stellar components is also referred to as semi-major axis $a$.  The COS's origin is placed at
the center between the two stellar components as previously done in Paper~I and II.
Observationally, the distance between the stellar components
is denoted\footnote{Occasionally, especially in observational astronomy, $a_{\rm bin}$
is referred to as semi-major axis as well, which is due to the different choices of the COS.}
as $a_{\rm bin}$, i.e.,
relationship
\begin{equation}
a \ = \ \frac{1}{2} a_{\rm bin} \ .
\end{equation}
From the perspective of orbital mechanics, it is of interest to consider cases of different
mass ratios $\mu$ given as
\begin{equation}
\mu \ = \ \frac{M_2}{M_1 + M_2}
\end{equation}
with $M_1$ and $M_2$ denoting the mass of the stellar primary (S1) and stellar secondary (S2),
respectively.  Clearly, both stellar components will orbit about the common center of mass (CM).
The distances of S1 and S2 from the CM are denoted as $Z$ and $a_{\rm orb}$, respectively
(see Fig.~4), with $a_{\rm orb}$ given as
\begin{equation}
a_{\rm orb} \ = \ 2a (1-{\mu})
\end{equation}
For equal-mass systems, i.e., $\mu = 1/2$, we find $a_{\rm orb}=a$, whereas in the limiting
case of very small mass for the secondary (e.g., the Sun-Jupiter pair, if treated as a binary),
we find $a_{\rm orb}~{\simeq}~2a$.  Information for $a_{\rm orb}/a$ values for systems of different
stellar mass ratios, including those considered here, is given in Figure~4.


\section{RESULTS AND DISCUSSIONS}


\subsection{Solution Landscapes}



Next we explore the general solution landscape for HZs of stellar binary systems.
Additional results pertaining to the widths of HZs, both with respect to the RVEM and GHZ,
can be found in Section~3.2.  Here we present a limited number of examples,
encompassing both equal-mass and non-equal-mass systems, while assuming either $e_{\rm bin} = 0.0$
or $e_{\rm bin} = 0.25$.  We also discuss the structure and extent of the GHZ for
Mars-type (0.1~$M_\oplus$), Earth-type (1.0~$M_\oplus$), and super-Earth-type planets
(5.0~$M_\oplus$).  Furthermore, we comment on the relationship between the respective
RHZ and the orbital stability limit for possible system planets.  Results for {\it P-}type HZs
are given in Figures~5 and 6, whereas results for {\it S-}type HZs are given in Figure~7.
Additional results are given in Tables 5 and 6.  Generally, orbital stability limits of possible
system planets may impede on the zones of habitability.  This type of behavior may occur
regarding {\it P-}type and {\it S-}type HZs.  If it does occur, {\it PT-}type and {\it ST-}type
HZs are obtained, respectively.

Regarding {\it P-}type HZs, our results, chosen in lieu of similar cases, read as follows:
For systems with binary semi-major axes $a_{\rm bin} = 0.25$~au, {\it P}-type GHZs can be found for both the
$M_1=M_2=1.00~M_{\odot}$ and the $M_1=1.00~M_{\odot}$ \& $M_2=0.50~M_{\odot}$ cases.
The outer limit for the equal-mass case ranges\footnote{For the GHZ and RVEM-type limits, as well as the
planetary orbital stability limits, values of higher precision than those given here are readily included
in Tables 5 to 10.  This was done mostly for tutorial reasons; it is known, however, that the
limits of stellar HZs and those of orbital stability are inherently uncertain owing to various processes
and effects not included in this study.  Examples include the particulars of atmospheric compositions,
space and planetary weather patterns, and tidal heating; see, e.g., \cite{ramir18} for details.
} 
from 2.37 to 2.38~au; thus, the minimum value which is 2.37~au must be taken as outer limit for the RHZ
(see Paper~I for further discussion).  For a Mars-type planet, the GHZ's inner limit ranges from 1.42 to 1.44~au,
for an Earth-type planet, it ranges from 1.34 to 1.36~au, and for a super-Earth-type planet, it ranges from 1.29 to 1.32~au.
Here the maximum values must be taken as adequate limits for the RHZs, given as 1.44, 1.36, and 1.32~au, respectively.
The orbital stability limit for possible system planets is given as 0.60~au.  It is located well inside
the inner limit of the RHZ for a Mars-type, Earth-type, or super-Earth-type planet.  Therefore, with respect
to the GHZ, the full extent of the RHZ is available in each case.

Figure~5 depicts the case of $M_1=1.00~M_{\odot}$ \& $M_2=0.50~M_{\odot}$ with $a_{\rm bin} = 0.25$~au.
Here both the outer and inner limits of the RHZs are reduced owing to the lower luminosity of the secondary
stellar component.  Here the outer limit of the RHZ is given as 1.65~au, whereas the inner limits for the RHZ
are given as 1.10, 1.05, and 1.01~au for a Mars-type, Earth-type, and super-Earth-type planet, respectively.
The orbital stability limit for possible system planets is given as 0.60~au.  It is again located well inside
the inner limit of the RHZs for the three cases studied.  Therefore, the full extent of the RHZ is again
available as GHZ in each case.

In case of $e_{\rm bin} = 0.25$, there are
only minor changes to the systems.  The outer limits of the RHZs are reduced, whereas the inner limits of the
RHZs are increased.  Additionally, the orbital stability limits for possible system planets, both for
the equal-mass and the non-equal-mass system, are increased.  However, in both cases, the orbital stability limits
remain below the limits of the RHZs; therefore the full extents of the RHZs are available as GHZ in each case.
However, the situation is considerably changed if systems of larger separations are considered.  Table~5 and 6
show some values for the habitability classification, including $e_{\rm bin}$ = 0.50 and 0.75.  In summary, the maximum
value of $a_{\rm bin}$ for {\it P-} or {\it PT-}type habitability to exist significantly decreases if
$e_{\rm bin}$ is increased.  Thus, there will be many systems without {\it P/PT-}type habitability, at least
following the approach adopted here.

This result is also exemplified by Figure~6, which is based on $a_{\rm bin}$ = 0.50~au.  In this case,
for the equal-mass system, the RHZ's outer limit is given as 2.36~au.  Here the inner limits for the RHZ
are given as 1.48, 1.40, and 1.37~au for a Mars-type, Earth-type, and super-Earth-type planet, respectively.
Here the orbital stability limit for possible system planets is given as 1.19~au, and therefore does not
impede on the RHZs.  Again, the full extents of the RHZs are available for facilitating habitability regarding
the GHZ.  However, this situation is drastically different for the non-equal-mass system of $M_1=1.00~M_{\odot}$ \&
$M_2=0.50~M_{\odot}$, or if the orbital eccentricity is changed to $e_{\rm bin}$ = 0.25.

If we take the equal-mass system with $M_1=M_2=1.00~M_{\odot}$ with $e_{\rm bin}$ = 0.25 as an example, the
RHZ's outer limit is given as 2.35~au.  Here the inner limits for the RHZ are given as 1.52, 1.44, and 1.40~au
for a Mars-type, Earth-type, and super-Earth-type planet, respectively, whereas the orbital stability limit
for possible system planets is given as 1.53~au.  Hence, the orbital stability limit is inside the RHZ for
all three cases of possible system planets, and the available HZ is notably reduced (i.e., {\it PT-}type
habitability).  Furthermore, in this type of system, the available HZ does not depend on the type of
system planet.
A more drastic case is given by the non-equal-mass system with $M_1=1.00~M_{\odot}$ \& $M_2=0.50~M_{\odot}$, 
$a_{\rm bin}$ = 0.50~au, and $e_{\rm bin}$ = 0.25.  Here the outer limit of the RHZ varies between 1.57 and
1.91~au, with 1.57~au identified as relevant number.  Moreover, the orbital stability limit for possible system
planets is given as 1.61~au.  Therefore, according to the model adopted in this work, there is no GHZ available
in this system.

{\it S-}type habitability always occurs if the distance between the stellar components is chosen sufficiently
large, noting that the minimum distance increases as a function of the stellar luminosities.  Figure~7 explores
some examples of {\it S-}type cases, taking the equal-mass and non-equal-mass systems, as previously discussed,
while assuming $a_{\rm bin}$ = 5.0~au as well as $e_{\rm bin}$ = 0.0 or 0.25.  Our calculations show that an
{\it S-}type GHZ is able to exist around the primary star for the non-equal mass case considered here.
The RHZ's outer limit has a minimum of 1.68~au, which is smaller than 1.69~au, the stability limit for
possible system planets; here $e_{\rm bin}$ = 0.0 is assumed.  The inner limits of the RHZ are 1.01, 0.95,
and 0.92~au for a Mars-type, Earth-type, and super-Earth-type planet, respectively.  Furthermore, an
{\it ST}-type GHZ is found around the stellar primary for the equal-mass case. The orbital stability limit,
given as 1.37~au, is notably smaller than the RHZ's outer limit, given as 1.73~au; it thus represents
the outer limit of GHZ.  Here the inner limits pertaining to the Mars-type, Earth-type and super-Earth-type
planets are given as 1.04, 0.98, and 0.94~au, respectively.

If larger eccentricities are chosen, changes occur, as expected, but they are much less profound than in
the {\it P-}type case.  For $e_{\rm bin}$ = 0.25, the changes in the RHZ limits are very small compared to
circular orbit in the non-equal mass system due to the reduced luminosity of the secondary as well as the large
separation distance.  A major difference, however, arises from the orbital stability limit, which shapes the
GHZ, as the outer limit is now identified as 1.17~au.  In case of the equal mass system with $e_{\rm bin}$ = 0.25,
the outer limit of RHZ changes to 1.71~au.  The inner limits of RHZ are identified as 1.09, 1.01, and 0.97~au
for the Mars-type, Earth-type, and super-Earth-type planets, respectively.  However, no HZ could be 
found in this case, as the orbital stability limit at 0.96~au is less than the RHZ's inner limit.

Table~5 and 6 list the limits for {\it P-}type, {\it PT-}type, {\it S-}type, and {\it ST-}type habitability
for different systems regarding both the GHZ and RVEM HZ.  It is revealed that higher eccentricities for the
system components always entail higher limiting values for $a_{\rm bin}$ pertaining to the
{\it S/ST-}type case.  High values of eccentricity mean that the two stellar components
will have relatively close approaches at periapsis implying increased impediment, as indicated by the structure
of the RHZs and orbital stability limits.  Therefore, for highly eccentric stellar orbits, relatively large
separation distances are required for {\it S-}type habitability to be realized.  Moreover, increased values
of eccentricity also adversely impact {\it P/PT-}type habitability, but the impact is less severe.


\subsection{Widths of Binary Habitable Zones}

\subsubsection{{\it P-}Type Case Studies}



Next we discuss some examples pertaining to the widths of binary HZs.  First, we focus on
{\it P-}type case studies.  Results are given in Figures 8, 9, and 10, as well as in Table 7 and 8.
The widths are given as a function of eccentricity
of the stellar components.  Combining the constraints (see Section 2.1), which include (1) the existence of the
RHZ (radiative criterion) and (2) the orbital stability limit of possible system planets
(gravitational criterion), the width of {\it P/PT}-type HZs is given as
\begin{equation}
{\rm Width}~(P/PT) \ = \ {\rm RHZ}_{\rm out} - {\rm Max} \big( {\rm RHZ}_{\rm in} , a_{\rm cr} \big)
\end{equation}
where ${\rm RHZ}_{\rm in}$ and ${\rm RHZ}_{\rm out}$ denote the inner and outer limit of the RHZ,
respectively, and $a_{\rm cr}$ denotes the limit of orbital stability (which is a lower limit).  No {\it P/PT}-type
HZ is found if the width following Eq.~(9) is less than or equal to zero.

Figures 8, 9, and 10 showcase systems of stars of different masses, with combinations for $M_1$
and $M_2$ of 1.25, 1.00, 0.75, and 0.50~$M_\odot$ taking into account $a_{\rm bin}$ = 0.25~au
and 0.50~au.  Regarding the eccentricity of the systems, we assume $e_{\rm bin}$ to take values
between 0.0 and 0.80.  Note that both RVEM HZs and GHZs are considered.  Figure 8 and 9 refer
to Earth-type planets; note that a total of 12 combinations of stellar masses are taken into account.
In the following, we focus on a limited number of examples aimed at highlighting distinct trends.

For the {\it P}-type RVEM (see Figure 8), the combination of $M_1 = M_2 = 0.75~M_\odot$ with 
$a_{\rm bin}$ = 0.25~au entails {\it PT}-type HZs for eccentricities larger than 0.148.
For mass combinations of $M_1 = 1.00~M_\odot$ \& $M_2 = 0.50~M_\odot$, the limiting value for
the eccentricity is given as 0.348.  Furthermore, for mass combinations of $M_1 = 1.00~M_\odot$
\& $M_2 = 0.75~M_\odot$, the respective value for the eccentricity is 0.630.  The three other
cases depicted in Figure~8 entail {\it P}-type HZ when the binary eccentricities
are less than or equal to 0.80 (the upper limit considered).
If the semi-major axis is taken as 0.50~au, the combination of $M_1 = M_2 = 0.75~M_\odot$
indicates {\it PT}-type HZs for eccentricities less than or equal to 0.272.  This values reads
0.286 for mass combinations of $M_1 = 1.00~M_\odot$ \& $M_2 = 0.50~M_\odot$, and 0.714 for
for mass combinations of $M_1 = 1.00~M_\odot$ \& $M_2 = 0.75~M_\odot$.  Cases of
{\it PT-}type HZs are found for $M_1 = M_2 = 1.00~M_\odot$ for all eccentricities
studied.  If the eccentricity is below 0.021, {\it P}-type HZs exist
for $M_1 = 1.25~M_\odot$ \& $M_2 = 0.50~M_\odot$.  For mass combinations of $M_1 = 1.25~M_\odot$
\& $M_2 = 0.75~M_\odot$, the limiting eccentricity is identified as 0.058.

Comparing the widths of HZs for the GHZ and different planetary climate models (see Figure~9),
the {\it PT}-type HZs correspond to smaller differences than {\it P}-type HZs.  This result is
as expected, noting that the stellar fluxes of the maximum greenhouse limit from the two models
are relatively similar, with both inner limits being set by the orbital stability limit.
As only the runaway greenhouse limit depends on the planetary mass, thus potentially affecting the
widths of their HZs, Earth-type planets are found to
have HZs of smaller widths than super-Earth-type planets, but their HZs are of larger widths than
for Mars-type planets in {\it P}-type cases.  For {\it PT}-type cases, those widths are always the same;
see Figure~10 for details.  Generally speaking, for fixed combinations of stellar masses, an increase
in eccentricity reduces the width of {\it P/PT}-type HZs in each case.  This effect is, however, less
pronounced if the separation distance between the stellar binary components is relatively small.

Table 7 depict selected critical values of eccentricities for different stellar mass combinations,
i.e., {\it P/PT}-type HZs are possible below those values but not at higher eccentricities.  Table~7
lists information for $a_{\rm bin}$ of 0.25~au and 0.50~au both with respect to the RVEM and GHZ; for
the latter, Earth-type planets are assumed.  For $a_{\rm bin} = 0.25$~au and the GHZ, 9 out of the 10
stellar mass combinations feature either {\it P-}type HZs or {\it PT-}type HZs, although for some
low-mass combinations there is a restriction regarding eccentricity.  However, this number is considerably
reduced for $a_{\rm bin} = 0.50$~au or if the RVEM-type HZ is selected.  Another finding is, however, that
one of the mass combinations falls short.  For equal-mass systems of $M_1 = M_2 = 0.50$ $M_\odot$ very
small separation distances need to be assumed in order to identify {\it P-}type HZs or {\it PT-}type HZs.
For the GHZ and Earth-type planets, that semi-major axis reads 0.124, 0.096, 0.082, and 0.074~au
for {\it P-}type HZs and 0.218, 0.212, 0.206, and 0.200~au for {\it PT-}type HZs with respect to
eccentricities of 0.00, 0.25, 0.50, and 0.75, respectively.  Table~8 depicts comparisons between
Mars-type, Earth-type, and super-Earth-type planets with respect to the GHZ.  No notable differences
are found with respect to high-mass combinations, especially concerning small binary separation distances,
but for low-mass combinations, planets of lower mass are advantageous for the existence of {\it P-}type HZs.
However, the orbital stability limit plays a key role in the existence of {\it P/PT-}type HZs.


\subsubsection{{\it S-}Type Case Studies}



In the following, we focus on {\it S-}type case studies.  Results are given in Figures 11, 12, and 13,
as well as in Tables 9 and 10.  The widths are given as a function of eccentricity
of the stellar components.  Combining the constraints (see Section 2.1), which include (1) the existence of the
RHZ (radiative criterion) and (2) the orbital stability limit of possible system planets
(gravitational criterion), the width of {\it S/ST}-type HZs is given as
\begin{equation}
{\rm Width}~(S/ST) \ = \  {\rm Min} \big( {\rm RHZ}_{\rm out} , a_{\rm cr} \big) - {\rm RHZ}_{\rm in}
\end{equation}
where ${\rm RHZ}_{\rm in}$ and ${\rm RHZ}_{\rm out}$ denote the inner and outer limit of the RHZ,
respectively, and $a_{\rm cr}$ denotes the limit of orbital stability (which is an upper limit).
No {\it S/ST}-type HZ is found if the width following Eq.~(10) is less than or equal to zero.

In the following, we focus on the study of {\it S}-type RVEM HZs and GHZs based on binary
semi-major axes of $a_{\rm bin} = 10.0$~au and 20.0~au.  Our aim is to identify the
maximum (i.e., critical) values of the eccentricity for the existence of the HZs.  In selected cases,
we also explore the differences in the HZs with respect to Mars-type, Earth-type, and super-Earth-type
planets.  We will focus on the same stellar mass combinations as done for {\it P-}type cases,
see Section 3.2.1.

Considering {\it S}-type RVEM HZs (see Figure 11 and Table 9) for equal-mass systems of $a_{\rm bin} = 10.0$~au
and $M_1 = M_2 = 1.00~M_\odot$, $M_1 = M_2 = 0.75~M_\odot$, and $M_1 = M_2 = 0.50~M_\odot$
the maximum eccentricities for {\it S}-type HZs to be able to exist are 0.295, 0.517, and 0.785, respectively.
The corresponding values for {\it ST}-type HZ are 0.642, 0.754, and at least 0.80
(the maximum value investigated).
This means a general increase in the maximal permissible eccentricity for {\it S/ST}-type HZs to be able to exist
in decreasing order of stellar masses and luminosities.
If we consider $a_{\rm bin} = 20.0$~au instead, the maximum eccentricities for {\it S}-type HZs to be able to exist
for $M_1 = M_2 = 1.00~M_\odot$ and $M_1 = M_2 = 0.75~M_\odot$ are given as 0.602 and 0.723, respectively.
The corresponding values for {\it ST}-type HZ are 0.791 for $M_1 = M_2 = 1.00~M_\odot$ and at least
0.80 for the other mass combinations considered in this study.

Furthermore, it is particularly interesting to explore the existence of {\it S}-type
and {\it ST}-type RVEM HZs for the non-equal-mass system of $M_1 = 1.25~M_\odot$ \& $M_2 = 0.75~M_\odot$.
Here the sum of the two masses is identical to the $M_1 = M_2 = 1.00~M_\odot$ system, but it has a notably higher
luminosity (see Table~4).  For this non-equal-mass system it is found that for
$a_{\rm bin} = 10.0$~au {\it S}-type and {\it ST}-type RVEM HZs ease to exist at eccentricities of
0.177 and 0.580, respectively.
Moreover, for $a_{\rm bin} = 20.0$~au {\it S}-type and {\it ST}-type RVEM HZs ease to exist at eccentricities of
0.529 and 0.755, respectively.
Hence, we are able to conclude that, in general, equal-mass systems are advantageous over non-equal-mass systems in
providing {\it S}-type and {\it ST}-type HZs for highly eccentric cases, and so are systems with relatively large separation distances.
Table~9 contains additional examples of non-equal-mass systems as, e.g., $M_1 = 1.00~M_\odot$ \& $M_2 = 0.50~M_\odot$,
which can be readily compared to the case of $M_1 = M_2 = 0.75~M_\odot$.  However, this does not necessarily indicate that the width of
the RVEM HZs is reduced compared to the corresponding equal-mass system, see Figure 11.

The differences in HZ widths can also be compared between different types of models.  In {\it S/ST}-type GHZs, the widths
are mostly determined by the differences in the stellar fluxes of runaway greenhouse limit (see Figure 12).
Note that the outer limit of the GHZ does not depend on the planetary mass.  Therefore, the HZ width of Earth-type planet
is always smaller than the one of super-Earth-type planet, and larger than the one that is Mars-type
(see Figure 13).  Table 9 and 10 depict selected critical values of eccentricities for different stellar mass combinations.
In almost all cases, the limiting values for the eccentricity is less than 0.80 to ensure the full availability of
{\it S/ST-}type habitability.  Again, it is found that values for the eccentricity, generally speaking, highest for
super-Earth-type planets, followed by Earth-type planets and Mars-type planets.
Generally, high values for the eccentricity affect the existence of HZs in {\it S-}type cases in an adverse manner.
In {\it S-}type cases, this phenomenon appears to be more pronounce than in {\it P-}type cases.  The adverse
effect of high eccentricities appears to be more notable regarding RVEM HZs, in considering of the fact
that those HZs are, generally speaking, more extended that GHZs.


\subsubsection{Comparative Studies based on Data by \cite{bar15}}


In the following, we present results of comparative studies between models involving the
data by \cite{man13} and \cite{bar15}.  We assume pairs of M-type dwarfs (equal-mass
cases) and pairs consisting of an M-dwarf and a star like the Sun (non-equal-mass
cases).  We consider models for {\it P-} and {\it S-}type habitability as well as
models involving either circular or elliptical orbits.  The stellar parameters of the
M-type dwarf are given in Table~11.  It shows that for the example as selected the
luminosity of the M-dwarf based on the data by \cite{bar15} is reduced by $\sim$20\% compared
to that based on the data by \cite{man13}.  Generally, the luminosities of M-dwarfs
are very small compared to solar-type stars.  Hence, the extents of HZs are significantly
reduced, and the positions of the HZs are much closer-in compared to stars of solar-type
luminosities.  The lower luminosity of the M-dwarf following \cite{bar15}
will further add to this outcome.

Table~12 conveys the results of our study for equal-mass M-type dwarfs.  As expected, the widths
of the HZs are reduced if the star data of \cite{bar15} are adopted, and those HZs are
positioned closer-in.  This behavior is identified for both {\it P-} and {\it S-}type models,
as well as for $e_{\rm bin}$ = 0.00 and 0.40, as explored.  However, those reductions are
minor only; they are identified as approximately 10\% for all cases studied.
Table~13 conveys the results for pairs containing an M-dwarf and a star like the Sun.
It is found that the differences between the models based on the \cite{man13} data
and the \cite{bar15} data are highly irrelevant, i.e., on the order of 0.1\% or
less.  In fact, the differences are totally insignificant for the kind of unequal pairs 
studied here as the more luminous star completely dominates the outcome.  If pairs with
lesser differences in the luminosities were considered, the expected
outcome would be intermediated between the cases of Table~12 and 13.


\section{SUMMARY AND CONCLUSIONS}

This studies extends the works previously presented in Paper~I and II.
The underlying mathematical concept follows a comprehensive approach for the computation
of HZs in stellar binary systems.  This approach is based on (1) the consideration of a
joint constraint including orbital stability and a habitable region
for a possible system planet through the stellar radiative energy fluxes
needs to be met; (2) the treatment of different types of HZs as
defined for the Solar System; the latter utilizes previous models of planetary
atmospheres around different types of main-sequence stars given by \cite{kop13,kop14};
(3) the provision of a combined
formalism for the assessment of both {\it S}-type and {\it P}-type HZs based on
detailed mathematical criteria, i.e., akin to previous work mathematical criteria
are presented for which kind of system {\it S}-type and {\it P}-type habitability
is realized; and (4) applications to stellar systems for different masses in either circular
or elliptical orbits.

With respect to the planetary climate models, we discuss both the GHZ and RVEM case
as defined here and previously used by other authors.  Regarding the outer limit for the GHZ,
we consider three different limits, pertaining to Mars-type, Earth-type, and super-Earth-type planets, entailing
somewhat different limits for the run-away greenhouse effect, as previously discussed by \cite{kop14}.  The
associated planetary masses are given as 0.1~$M_\oplus$, 1.0~$M_\oplus$, and 5.0~$M_\oplus$, respectively.
Our theoretical simulations focus on stars with fixed masses given as 1.25~$M_\odot$, 1.00~$M_\odot$, 0.75~$M_\odot$,
and 0.50~$M_\odot$, respectively.  Special attention should be given to pairs of low-mass stars, i.e., 0.75~$M_\odot$
and 0.50~$M_\odot$, considering that the frequency of those stars is significantly higher compared
to high-mass stars (including the Sun) as pointed out by, e.g., \cite{kro01,kro02}.  Note that the
astrophysical potential of those stars has been explored by, e.g., \cite{hel14}, \cite{kas14}, and \cite{cun16}.

Regarding the work given in Paper~I and II, we confirm the following findings:

\noindent
(1) As discussed in Paper~I and II, the solution of the underlying quartic equation
proves to be a powerful tool to identify {\it S-}type and {\it P-}type habitability
in stellar binary systems.  The cases of {\it ST-}type and {\it PT-}type habitability
can be identified as well.

\noindent
(2)  If the stellar system components have a sufficiently large separation, {\it S/ST-}type
habitability is identified.  If the separation is sufficiently small, {\it P/PT-}type
habitability is identified; this result is in alignment with previous findings given
in the literature.

\noindent
(3)  Generally, stellar components with relatively high luminosities favor large-widths HZs.
However, the HZs are significantly reduced in size (or even eliminated) if the luminosities
of the two stellar components are highly unequal.  This effect applies to {\it P/PT-}type HZs,
and it is mostly due to the reduction of the respective RHZs.

\noindent
(4) Concerning the facilitation of habitability, systems of low eccentricity are
advantageous compared to systems of high eccentricity.  This outcome is due to the
different structures of the RHZs as well as the fact that high eccentricities for the
stellar components have an adverse impact on planetary orbital stability.
Adverse impacts due to high eccentricities are most pronounced for {\it S/ST-}type HZs.
In addition, adverse impacts are most pronounced in systems of low-luminosities stars,
i.e., K and M dwarfs.

Our main findings concerning the work of this study includes:

\noindent
(1) The width of the HZs, indentified for the GHZ, is notably increased for
super-Earth-type planets, compared to Earth-type and Mars-type planets.  The reason
is that the GHZ's inner limit is set by the runaway greenhouse effect, see \cite{kop14},
which moderately depends on the planet's mass.

\noindent
(2) Exceptions regarding (1) are observed for {\it P-}type cases at high stellar eccentricities.
Here the width of the HZs is determined by the orbital stability requirement rather than the
inner limits of the RHZs.  Therefore, the general results about the widths of the HZs are
independent of the type of planet taken into account.

\noindent
(3)  Relatively large widths for the HZs (if existing) are identified for the GHZ in systems of
super-Earth-type planets compared to Earth-type planets, and moreover compared to Mars-type planets.
This statement particularly applies to systems where the eccentricity of the binary components
is small, the luminosities of the binary components are high, or both.

\noindent
(4) As default, we used the stellar data by \cite{man13}; however, we also pursued selected
comparative studies based on the data by \cite{bar15}.  The differences in the context of
this study are mostly insignificant, except that the latter data imply lower luminosities for
M dwarfs.  Thus, the widths of HZs are slightly reduced for pairs of those kinds of stars, but not
for systems containing a high-luminosity component as the impact of the latter is most decisive.

\noindent
(5) This study also employed updated limits for the HZs as given by \cite{kop13,kop14} compared
to limits previously available in the literature.  The most pronounced difference is found
for the GHZ that is significantly reduced.  This outcome reveals itself in calculations
for the widths of both {\it P-}type and {\it S-}type cases.

\noindent
(6) There are notable differences in the figures on the widths of HZs, displayed as function
of $e_{\rm bin}$, for {\it S}-type and {\it P}-type cases conveying the turnoff points;
the latter are given by the impact of planetary orbital stability.  In {\it S}-type cases,
the orbital stability limit acts as an outer limit for the HZs, which is relevant for
large values of $e_{\rm bin}$, truncating the GHZ for cases with different planetary masses at the same eccentricity.
For small values of $e_{\rm bin}$, the widths of the HZs are determined by the RHZs, and
thus with respect to the GHZ, there are different outcomes for Mars-type, Earth-type, and
super-Earth-type planets.  In {\it P}-type cases, the orbital stability limit acts as an
inner limit for the HZs, again relevant for large values of $e_{\rm bin}$, but that impact
is less pronounced.  Hence, there are once again differences between Mars-type, Earth-type,
and super-Earth-type planets occurring at low values of $e_{\rm bin}$.

Besides the results described here, there is a serious need for future work, including various theoretical expansions
as well as observational verifications with the help of existing and future space missions.  A crucial aspect pertains
to the consideration of alternate and expanded definitions of planetary habitability.  Examples of recent ideas
include work by \cite{pie11}, and \cite{ram18} who identified possible extensions of the classical HZ due to hydrogen and
methane clouds, respectively.  Furthermore, according to previous studies planetary habitability may be able to persist
if planets temporarily exit the stellar HZ, and re-enter.

This behavior is particularly relevant for planets with thick atmospheres.
In that context, a readily encountered scenario pertains to planets in highly elliptical orbits; see,
e.g., \cite{wil02}, \cite{kan12}, and \cite{pil16} for previous work.  In this case, owing to the ability
of the planetary atmosphere to store energy, the average (or median) amount of stellar radiation seems
to matter most rather than the orbit-dependent amounts.  Another important future application of our methodology,
i.e., the identification of {\it S-}type and {\it P-}type HZs through a joint formalism should include the exploration
of relationships between the existence of those zones and stellar evolution, i.e., the slow but persistent departure
of one of the stellar components from the main-sequence.  Previous studies outside of the methodology described
in this paper series have been given by, e.g., \cite{und03}, \cite{asg04}, \cite{jon05}, \cite{tru15}, and \cite{gal17}.
General aspects and open questions about planets that are also relevant for planets hosted by binary systems have been
discussed by, e.g., \cite{kas03}, \cite{lam09}, \cite{kal17}, \cite{ramir18}, among others.


\acknowledgments
The authors wish to draw the reader's attention to the online tool
{\tt BinHab 2.0}, hosted at The University of Texas at Arlington (UTA),
which allows the calculation of habitable regions in binary systems
based on the developed method.  This tool has been updated in January 2019.


\appendix

\section{GLOSSARY OF ACRONYMS}

\begin{tabular}{ll}
{\it Acronym}     & {\it Meaning} \\
\noalign{\bigskip}
\hline
\noalign{\bigskip}
     CHZ      &    Conservative Habitable Zone~(see Paper I and II)       \\
     COS      &    Coordinate System                                      \\
     EHZ      &    Extended Habitable Zone~(see Paper I and II)           \\
     GHZ      &    General Habitable Zone                                 \\
     HZ       &    Habitable Zone                                         \\            
     RHZ      &    Radiative Habitable Zone                               \\
     RVEM     &    Recent Venus / Early Mars (Zone)                       \\
\end{tabular}


\clearpage

\section{EFFECTIVE STELLAR FLUX APPROXIMATIONS}

In the following, we compare various effective stellar flux approximations available in the literature,
which are needed for calculating HZ limits for stars of different effective temperatures (see Figure~2).
Originally, \cite{kas93} provided stellar fluxes for standard M0, G2, and F0 main-sequence stars with
effective temperatures of 3700~K, 5700~K, and 7200~K, respectively.  In order to apply their climate model
to main-sequence stars of different spectral types, fitting equations are required.  Previously, the
two most widely used were those of \cite{und03} and \cite{sel07}.

\cite{und03} adopted a set of parabolic equations to fit the stellar fluxes, in units of solar constant,
dependent of the choice of $s_\ell$ (see Section~2.2), which are given as
\begin{equation}
S_{\rm eff} = {a_y}T_{\rm eff}^{2}+ {b_y}T_{\rm eff} + {c_y} \ .
\end{equation}
Here $T_{\rm eff}$ denotes the stellar effective temperature; see their text for coefficient data.
\cite{sel07} provided another formalism for interpolating the stellar fluxes as also used by \cite{cun14,cun15};
it is considered here as well for comparison.  The stellar fluxes, in units of solar constant, read
\begin{equation}
S_{\rm eff}=(s_{\ell}-a_{x}T_{**}-b_{x}T^{2}_{**})^{-2} \ .
\end{equation}
Here $s_{\ell}$ represents different HZ limits following \cite{kas93}, see Table~2, with
$T_{**} = T_{\rm eff}-5700$~K.  \cite{sel07} also conveys the various coefficients $a_x$ and $b_x$ for the
various climate models with boundaries tagged as $s_{\ell}$.

\cite{kop13} obtained new results for solar limits of habitability based on their updated climate model, which 
is a considerably revised version of the one previously given by \cite{kas93} and also used by \cite{wan17}.  
This updated 1D radiative convective, cloud-free climate model has new $\rm{CO_2}$ and $\rm{H_2O}$ 
absorption coefficients, taken from new data bases, a crucial improvement over previous works.  Moreover, 
they use a recast form for the equation of stellar flux calculation given as
\begin{equation}
S_{\rm eff}=S_{\rm eff \odot}+aT_*+bT^2_*+cT^3_*+dT^4_*
\end{equation}
Here $T_*=T_{\rm eff}-5780$ K; see their table~3 for information on the coefficients.

Subsequently, another revision was made by \cite{kop14} who also took the planetary mass into account, with
detailed consideration of N$_2$ background pressure.  That model considered Mars-type (0.1~$M_\oplus$),
Earth-type, and super-Earth-type (5.0~$M_\oplus$) planets --- as done as part of our study as well.
They found that the runaway greenhouse limit notably depends on the planetary mass.  A very weak dependency,
though negligible in consideration of the model's uncertainties, was also identified for the Early Mars
limit; thus it is ignored here as well.  Moreover, they also gave some minor updates for some
of the other coefficients previously obtained by \cite{kop13}; see their table~1 for information.



\clearpage


\begin{figure*} 
\centering
\begin{tabular}{c}
\epsfig{file=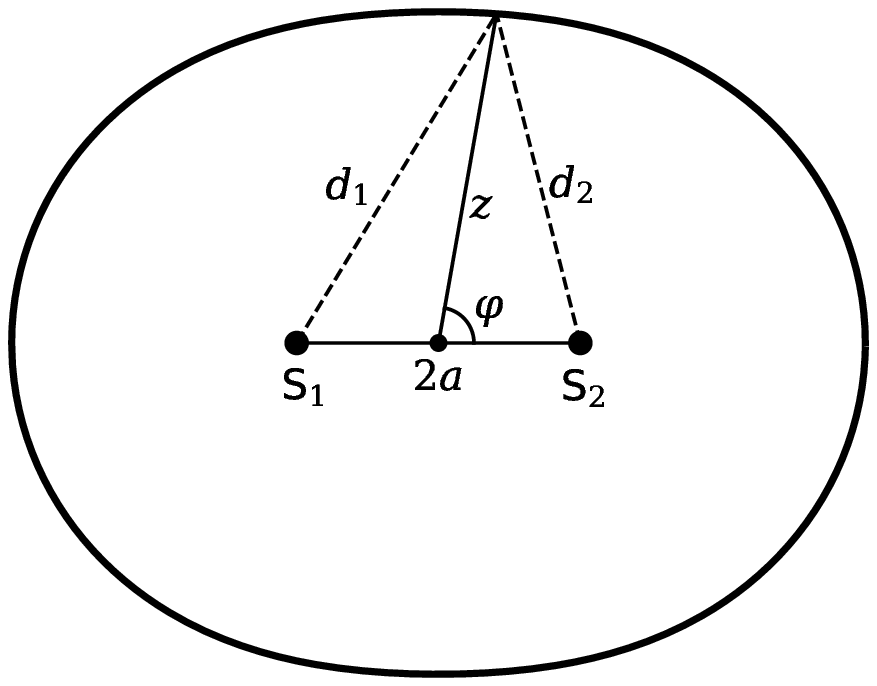,width=0.70\linewidth} \\
\noalign{\bigskip}
\noalign{\bigskip}
\epsfig{file=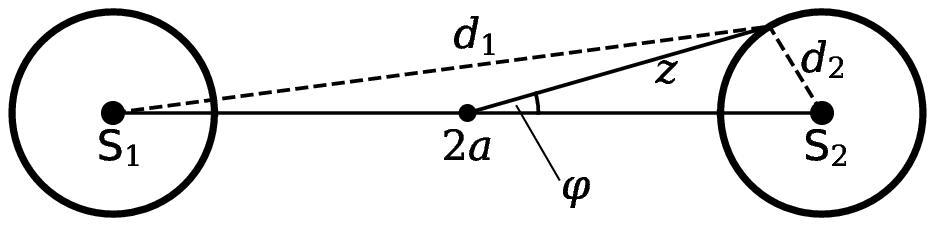,width=0.70\linewidth} \\
\end{tabular}
\caption{Set-up for the mathematical treatment of {\it P}-type (top)
and {\it S}-type (bottom) habitable zones of binary systems as given
by the stellar radiative fluxes with $a \equiv \frac{1}{2} a_{\rm bin}$.  Note that
the stars S1 and S2 have been depicted as identical for convenience.
}
\end{figure*}
\clearpage


\begin{figure*}
\centering
\begin{tabular}{cc}
\epsfig{file=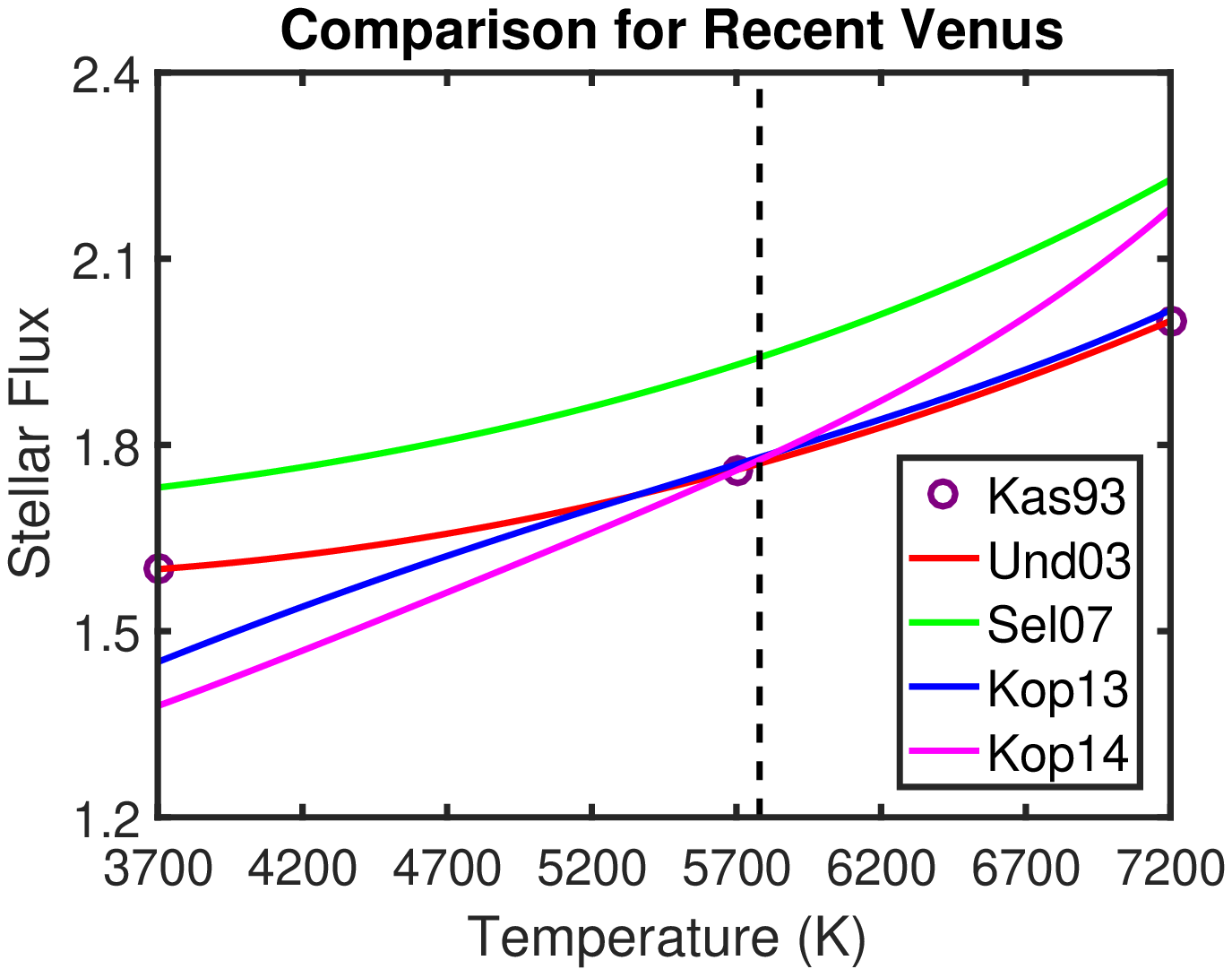,width=0.5\linewidth} &
\epsfig{file=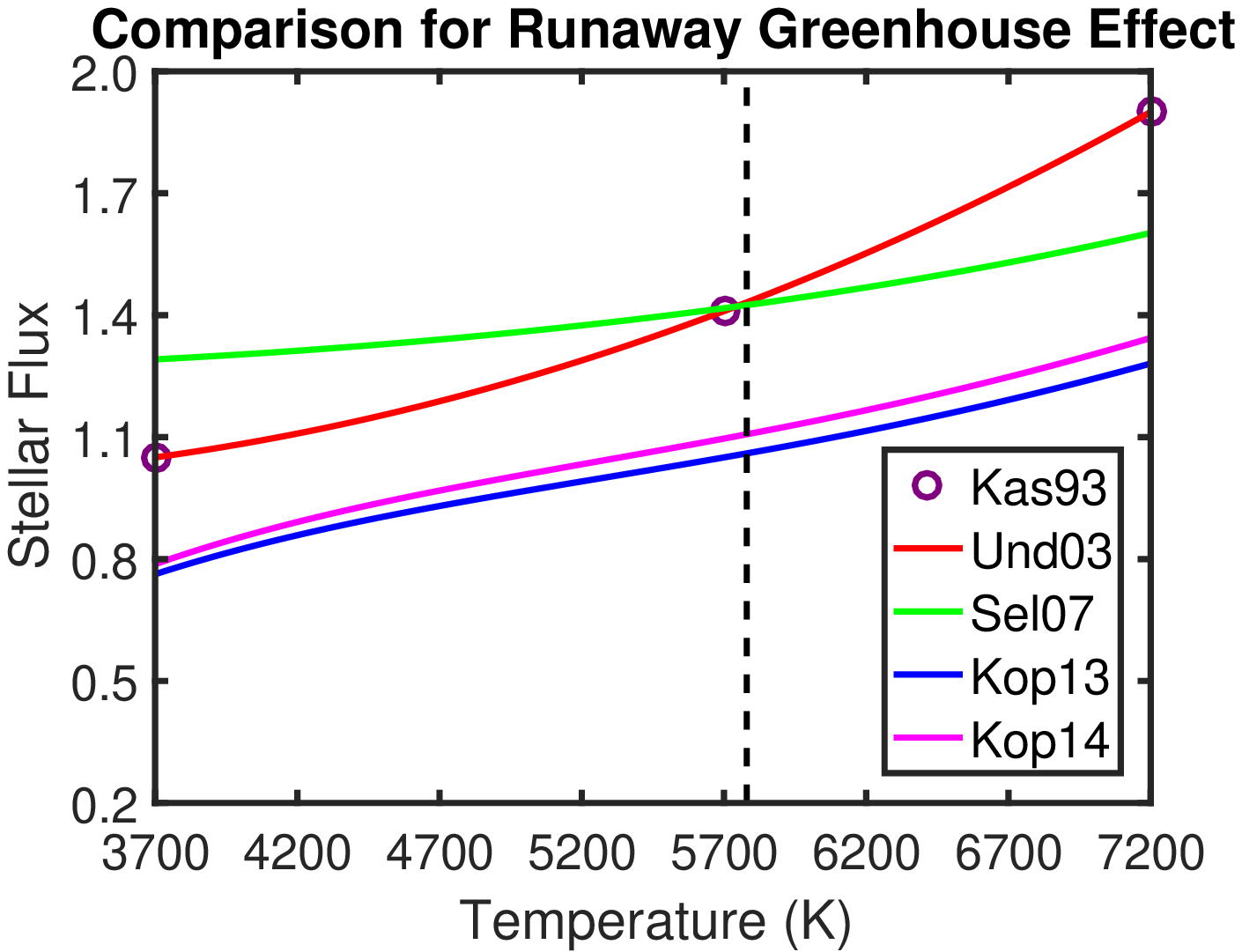,width=0.5\linewidth} \\
\epsfig{file=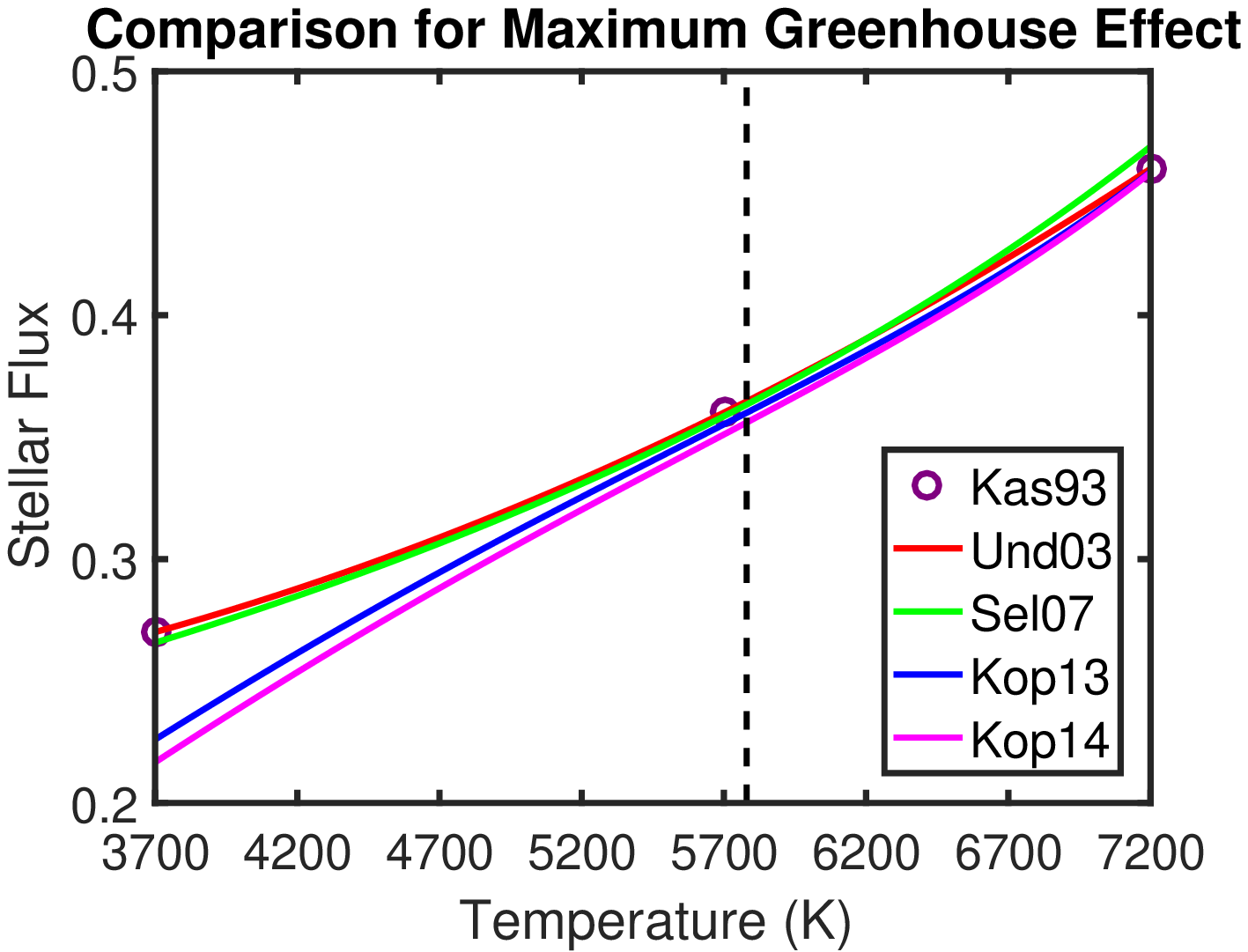,width=0.5\linewidth} &
\epsfig{file=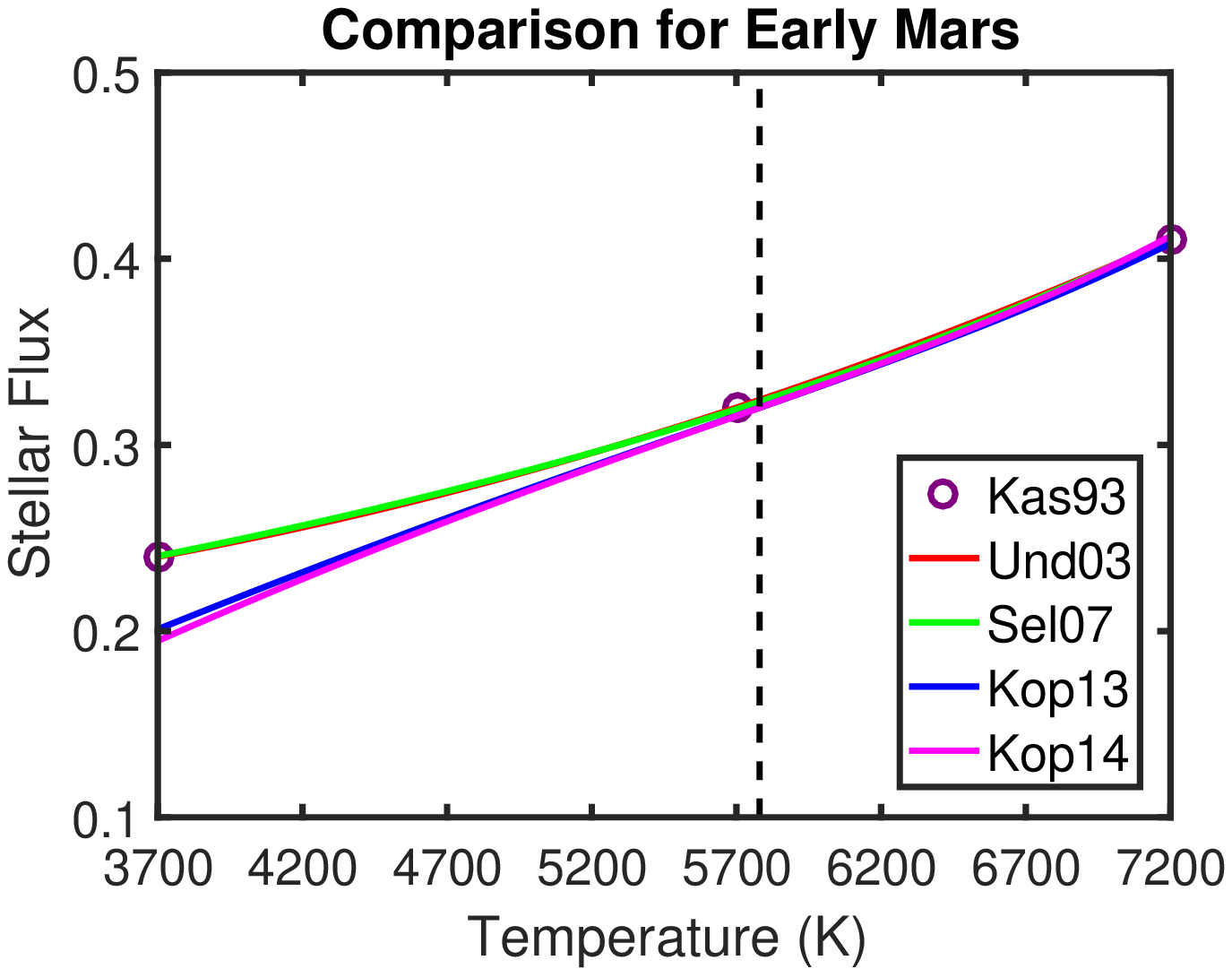,width=0.5\linewidth} \\
\end{tabular}
\caption{
Comparison of effective stellar flux approximations from different works.
Here the three circles indicate the stellar temperature data previously used by \cite{kas93},
whereas the dashed line depicts the now accepted value for the solar effective
temperature given as 5777~K.
}
\end{figure*}
\clearpage


\begin{figure*} 
\centering
\begin{tabular}{c}
\epsfig{file=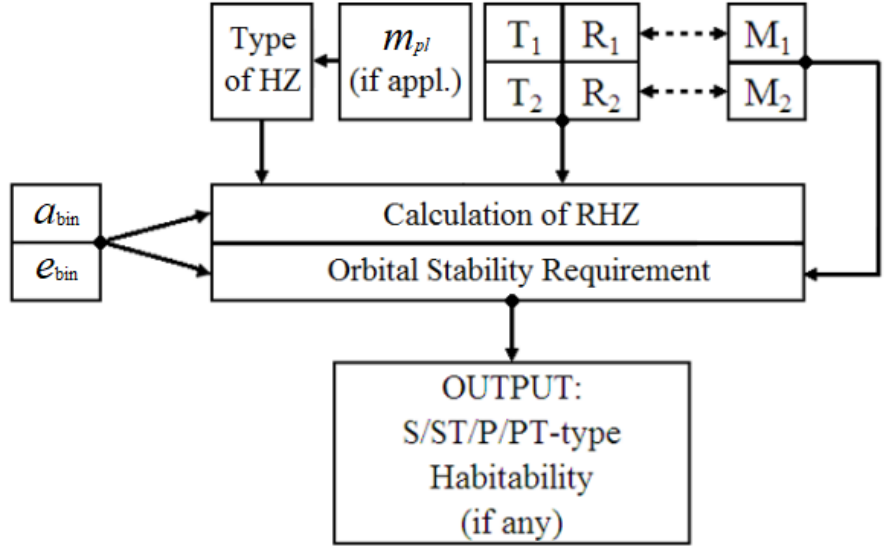,width=0.95\linewidth} \\
\end{tabular}
\caption{Flow diagram of {\tt BinHab 2.0} indicating the adopted method of solution.
Feed-ins and the generation of the output are indicated by arrowed solid lines.
For theoretical main-sequence stars, effective temperatures and stellar radii
($T_i$, $R_i$, with $i=1,2$), on the one hand, or stellar masses ($M_i$), on the
other hand, may serve as input parameters, as indicated by double-arrowed dashed
lines.  See \cite{cunb14} for information on the original version of {\tt BinHab}.
An updated version of that tool has been deployed in 2018, largely developed by one
of us (Zh. W.), which among other updates also takes into account the planetary mass
$m_{\rm pl}$.
}
\end{figure*}
\clearpage


\begin{figure*} 
\centering
\begin{tabular}{c}
\epsfig{file=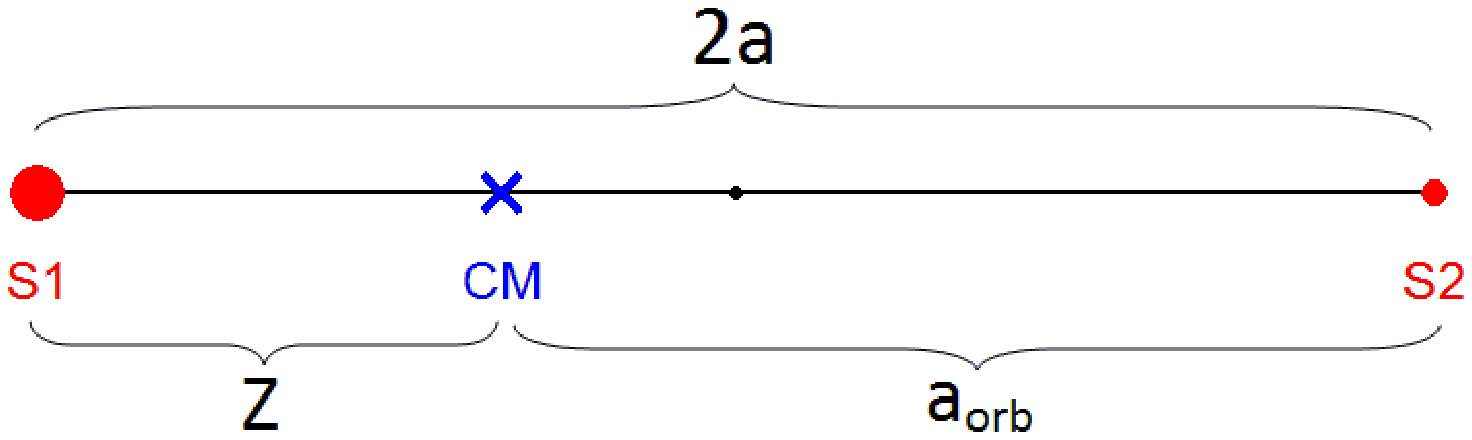,width=0.7\linewidth} \\
\epsfig{file=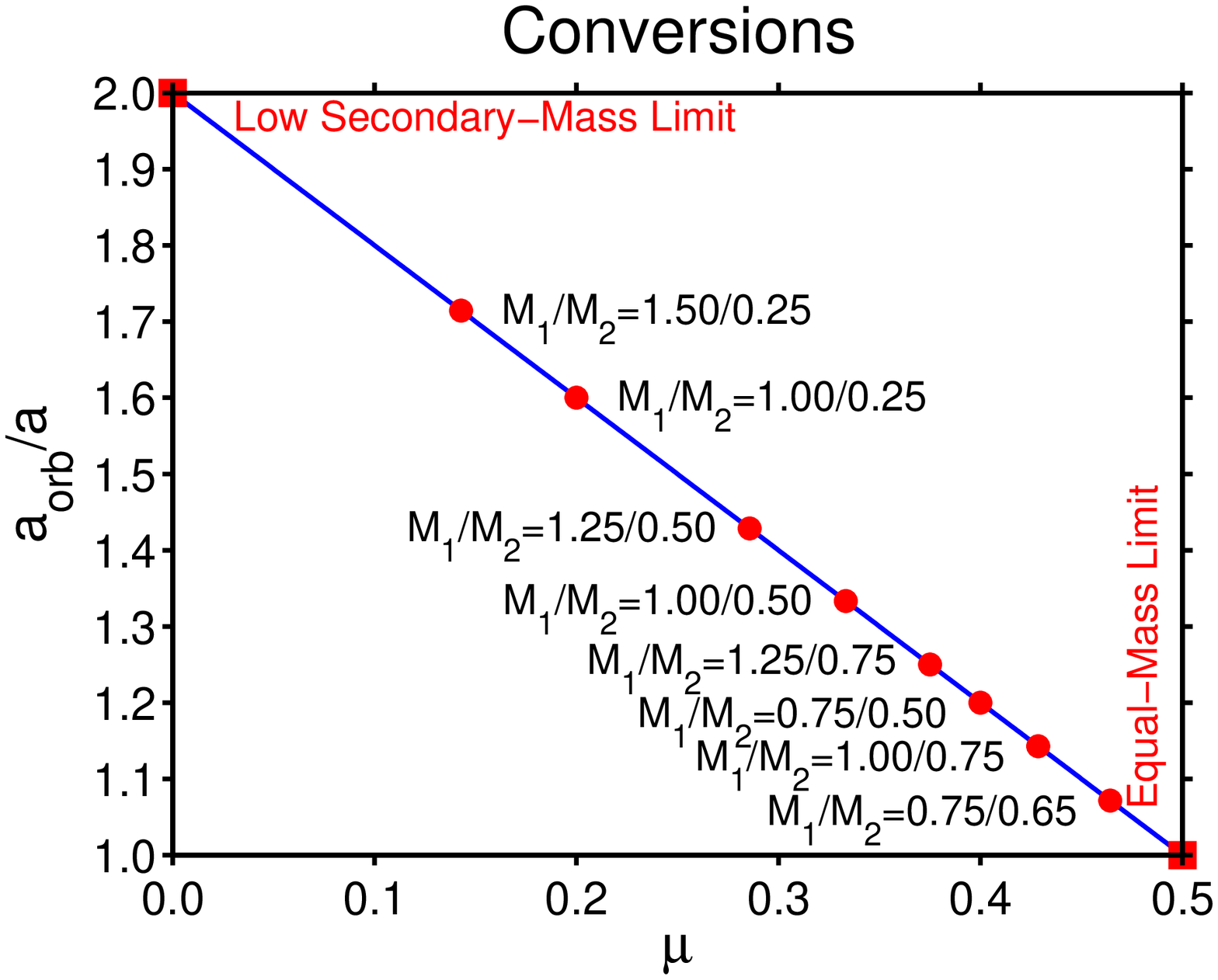,width=0.7\linewidth} \\
\end{tabular}
\caption{
Coordinate system aimed at defining $a_{\rm orb}$.  Here S1 and S2 denote
the two stellar components, which by default are of different masses.
CM denotes the center of mass. Ratios $a_{\rm orb}/a$ for different mass ratios 
$\mu$. The equal-mass limit $M_1=M_2$ and the low secondary-mass limit 
$M_2/M_1 \rightarrow 0$ are conveyed for convenience. 
}
\end{figure*}
\clearpage


\begin{figure*} 
\centering
\begin{tabular}{cc}
\epsfig{file=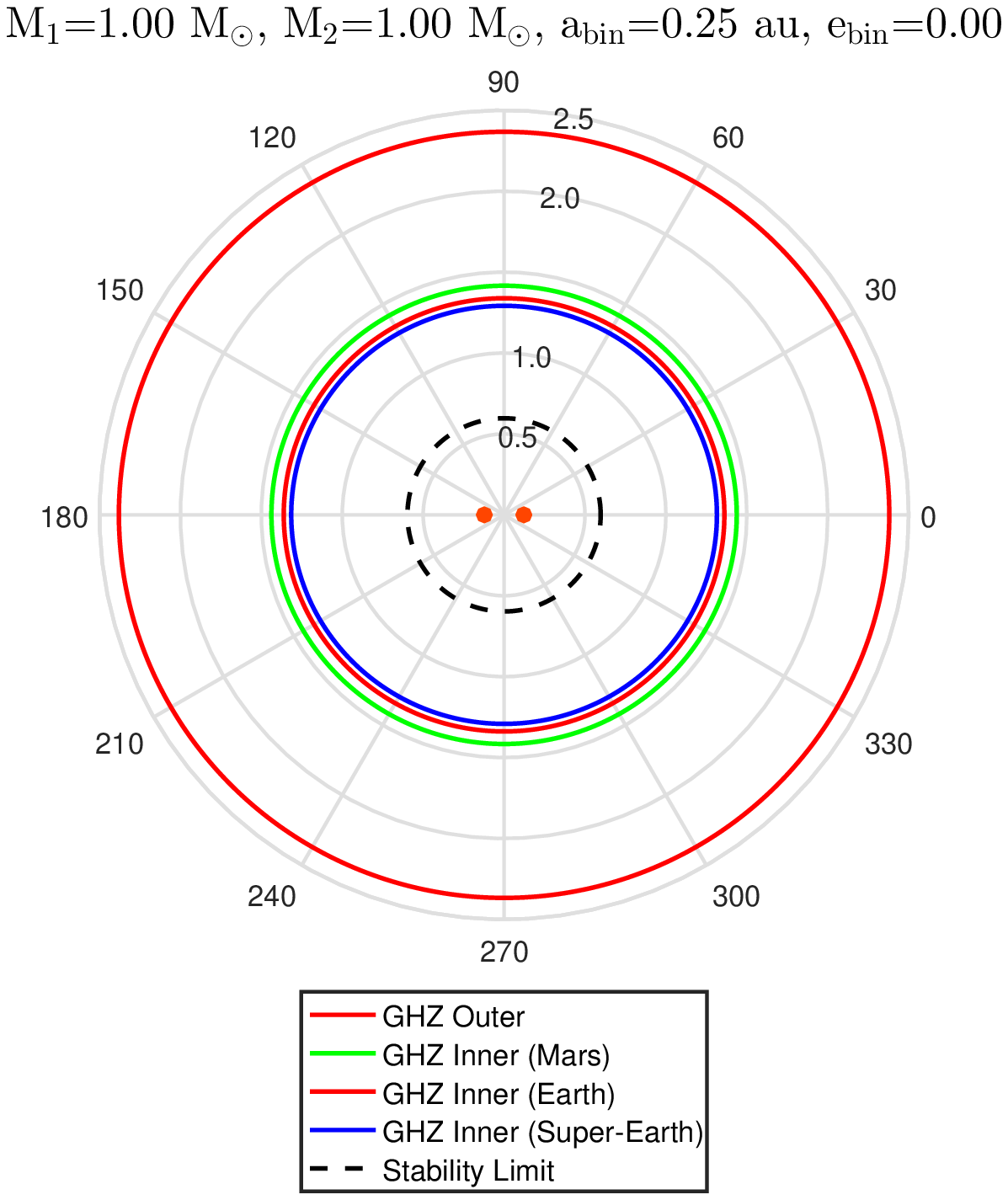,width=0.50\linewidth}
\epsfig{file=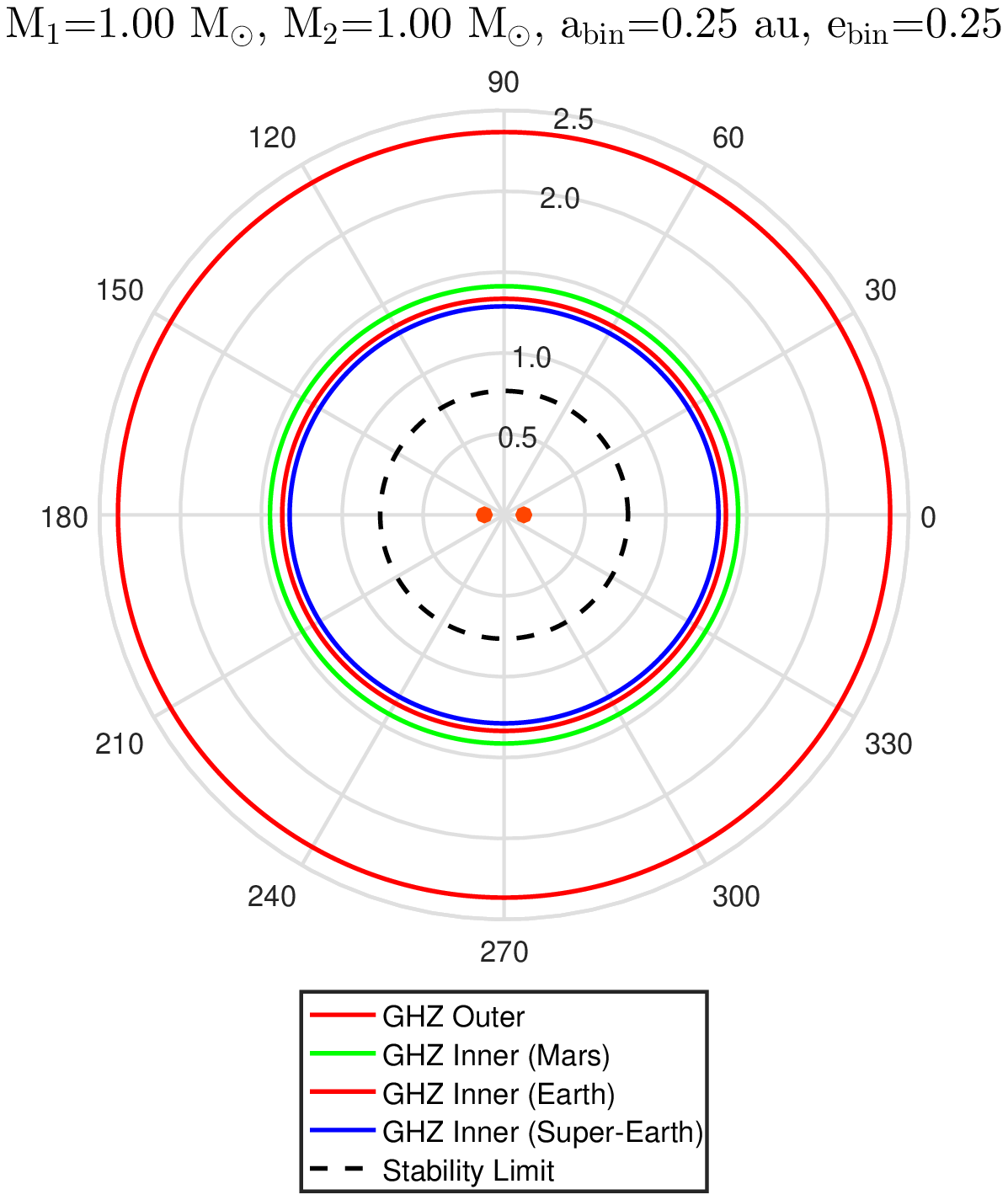,width=0.50\linewidth} \\
\epsfig{file=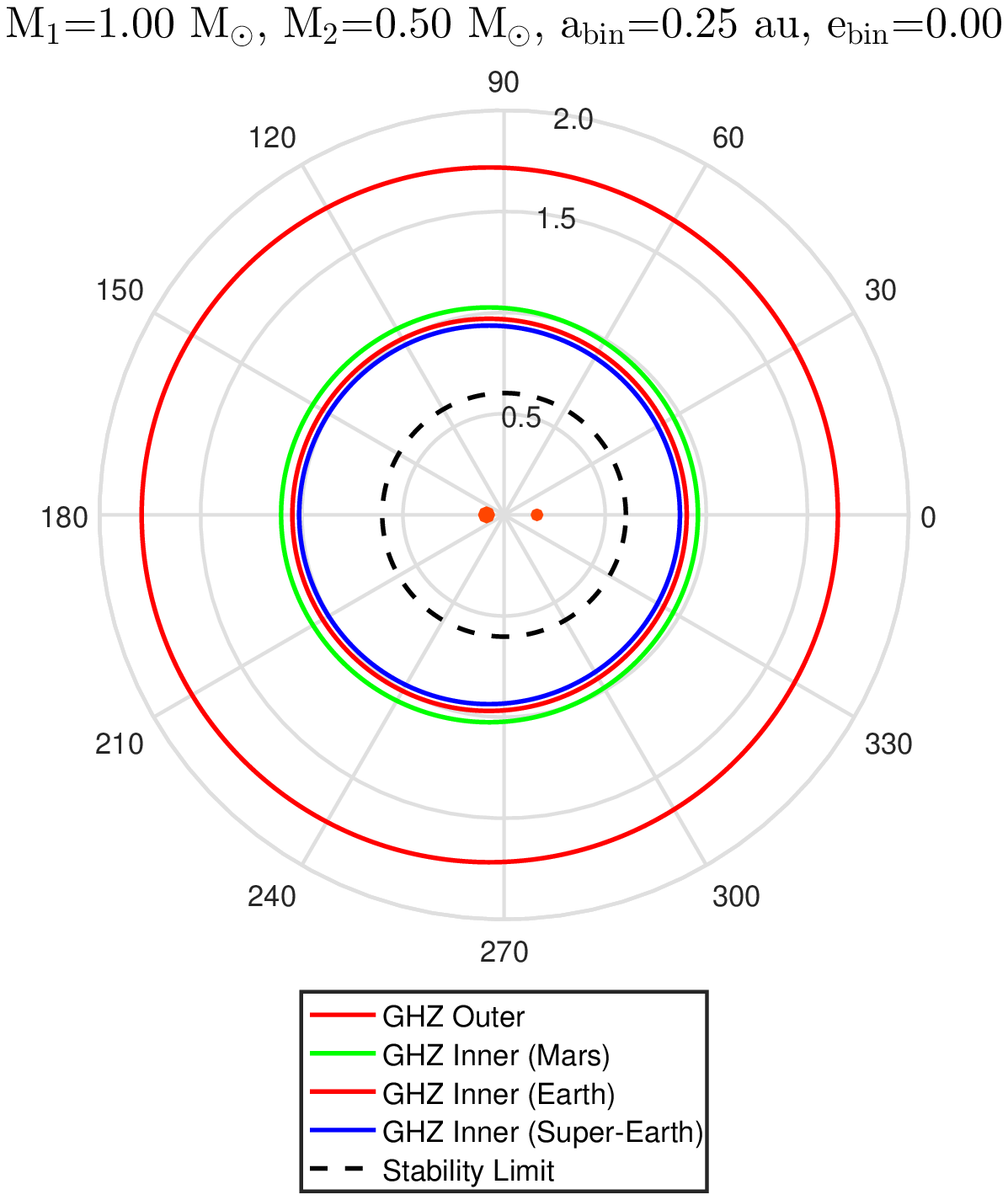,width=0.50\linewidth}
\epsfig{file=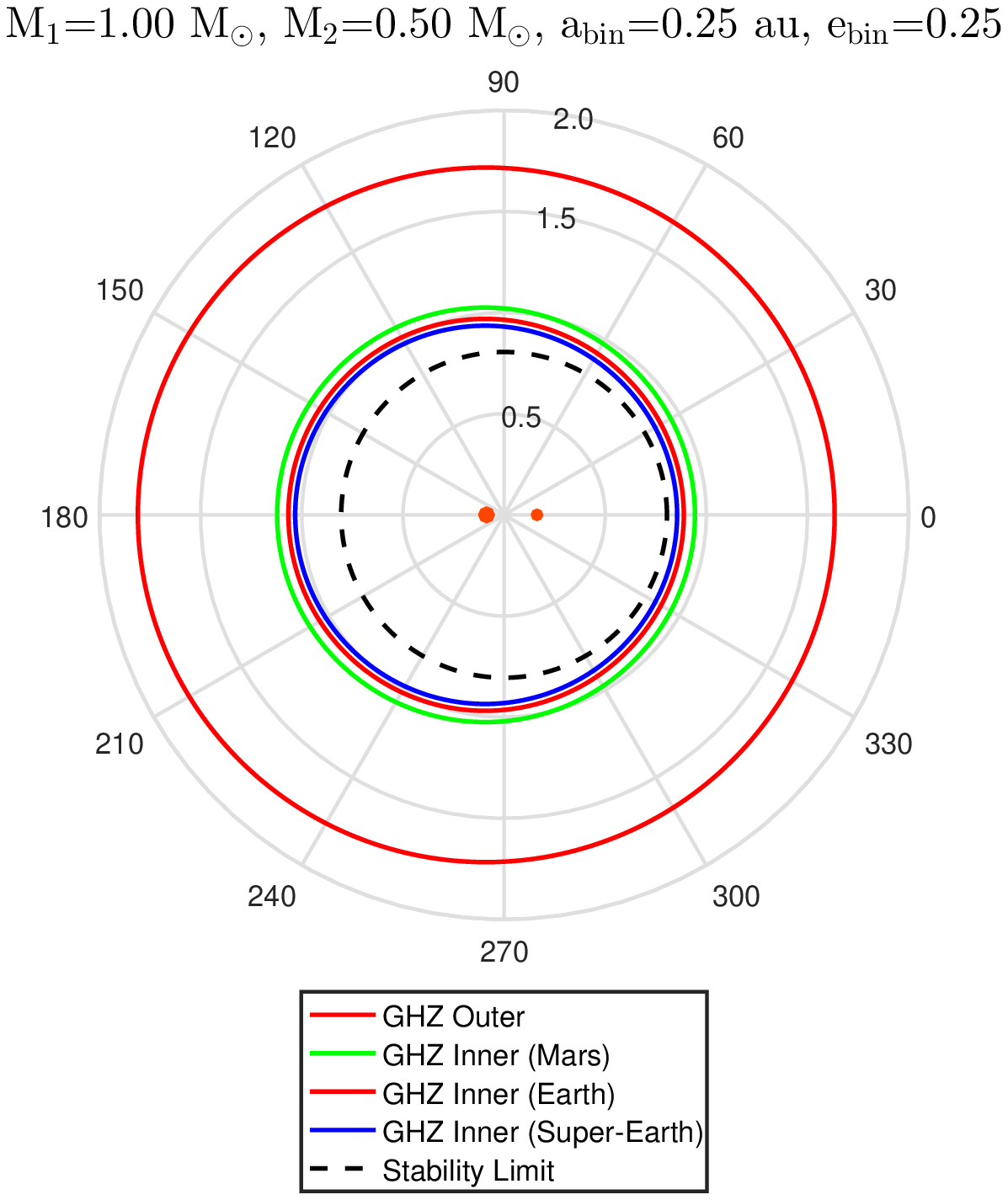,width=0.50\linewidth} \\
\end{tabular}
\caption{
Polar diagrams for {\it P}-type habitable zones for the GHZ, assuming Mars-type (0.1~$M_\oplus$), Earth-type,
and super-Earth-type (5.0~$M_\oplus$) planets, for two different combinations of stellar masses.
The stability limit (which is an inner limit) is given as well.
Here $a_{\rm bin}$ = 0.25~au is used.  Results are shown for $e_{\rm bin}$ = 0.00 and 0.25.
}
\end{figure*}
\clearpage


\begin{figure*} 
\centering
\begin{tabular}{cc}
\epsfig{file=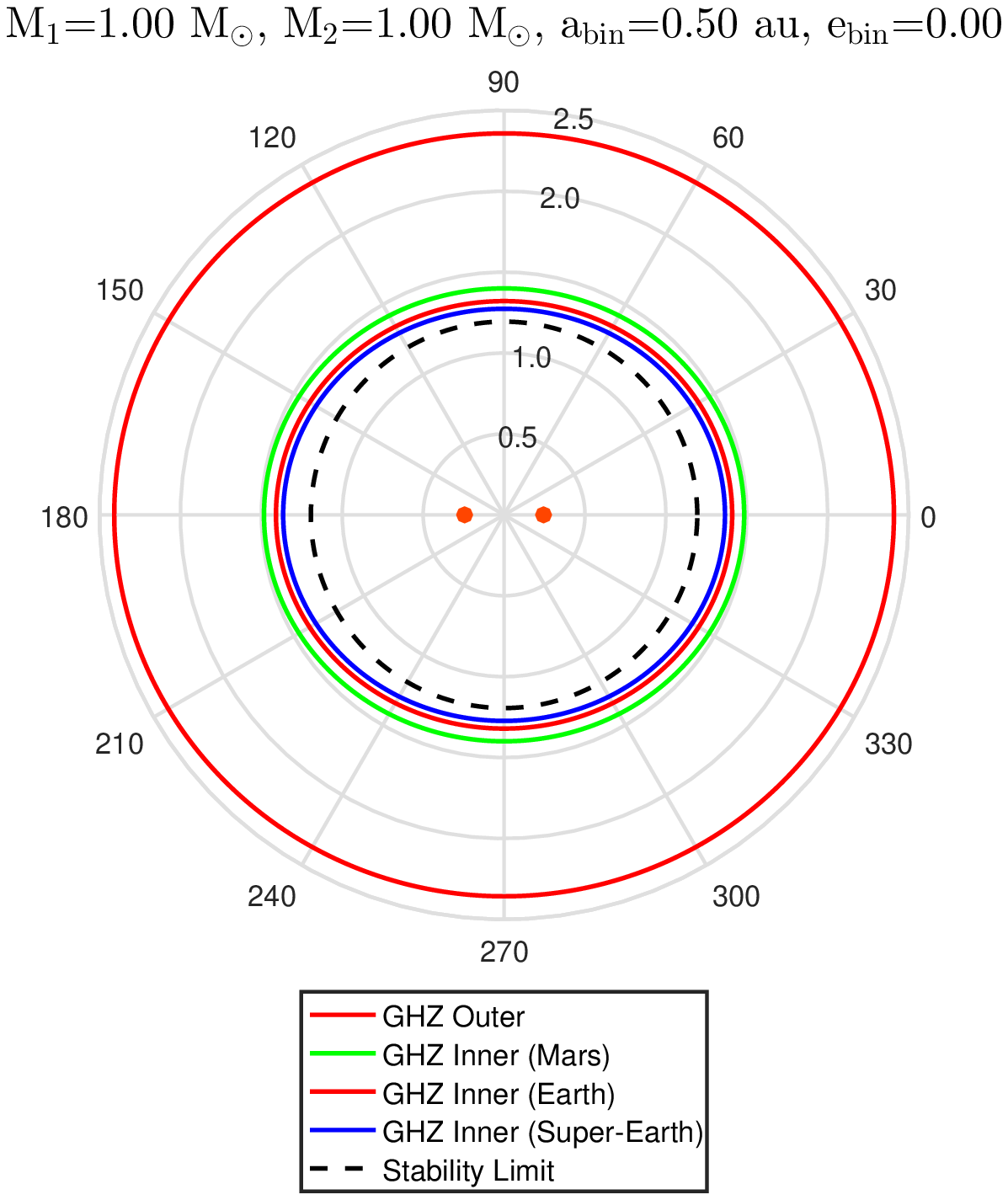,width=0.50\linewidth}
\epsfig{file=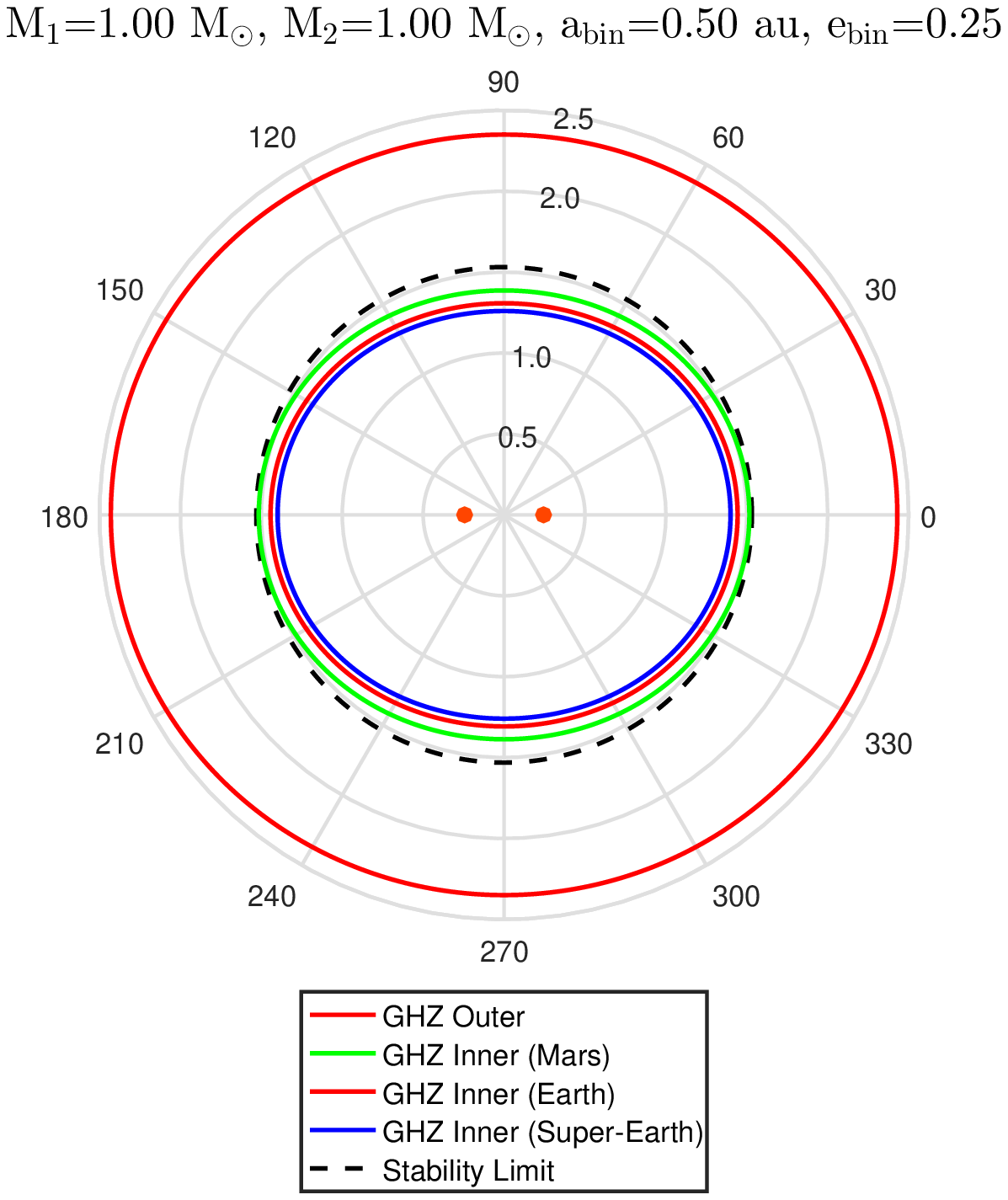,width=0.50\linewidth} \\
\epsfig{file=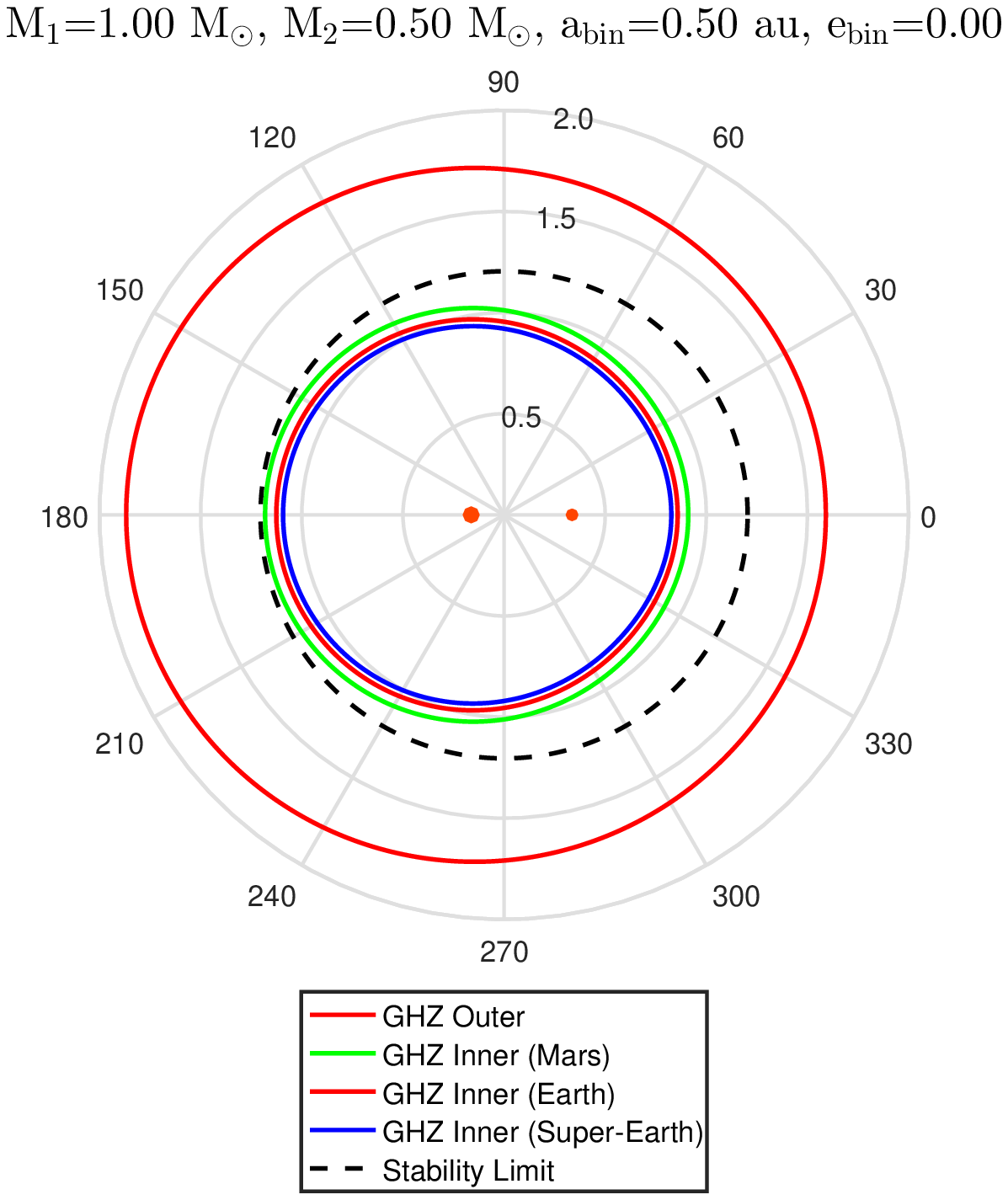,width=0.50\linewidth}
\epsfig{file=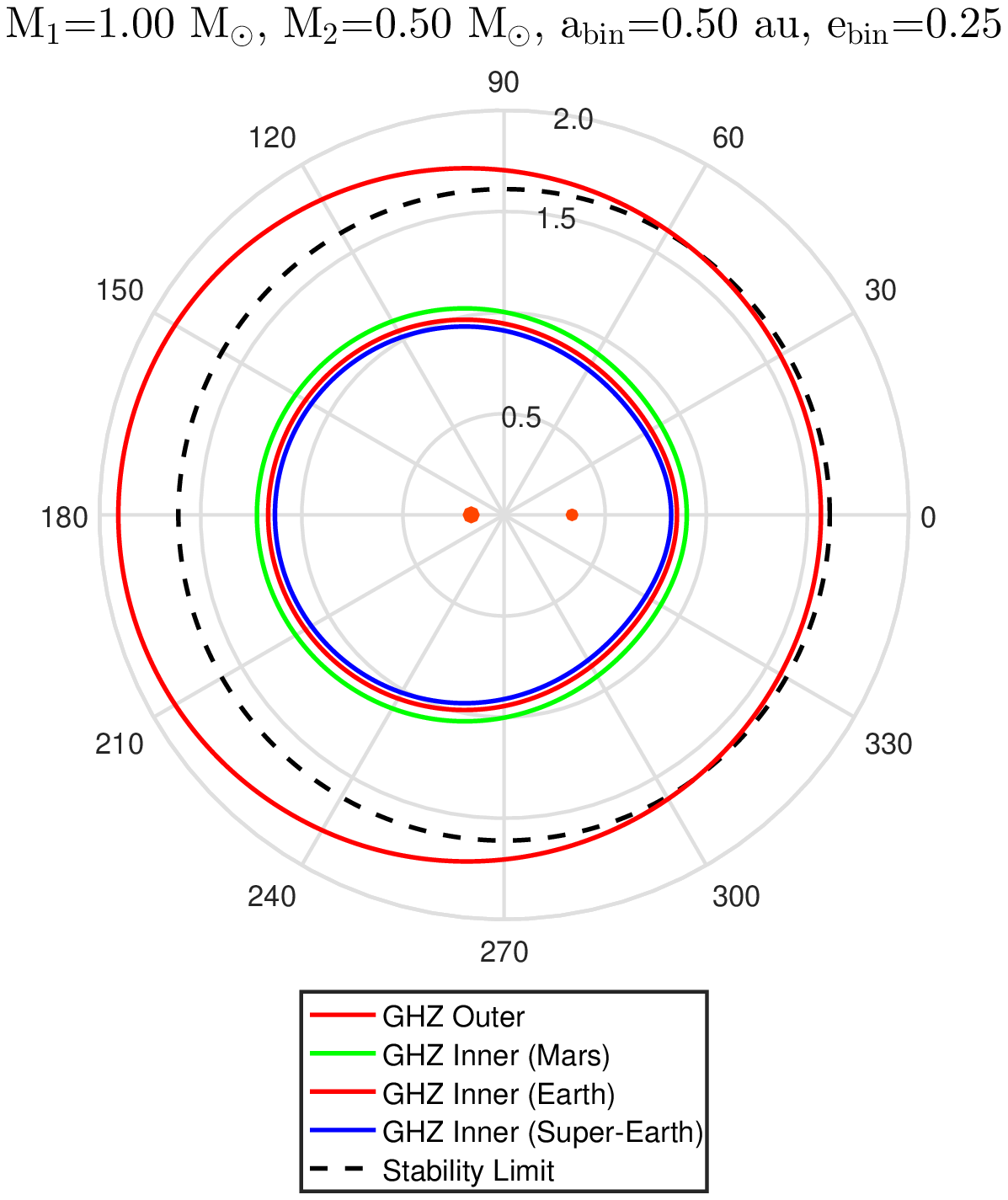,width=0.50\linewidth} \\
\end{tabular}
\caption{
Same as Fig.~5, but with $a_{\rm bin}$ = 0.50~au.
}
\end{figure*}
\clearpage


\begin{figure*} 
\centering
\begin{tabular}{cc}
\epsfig{file=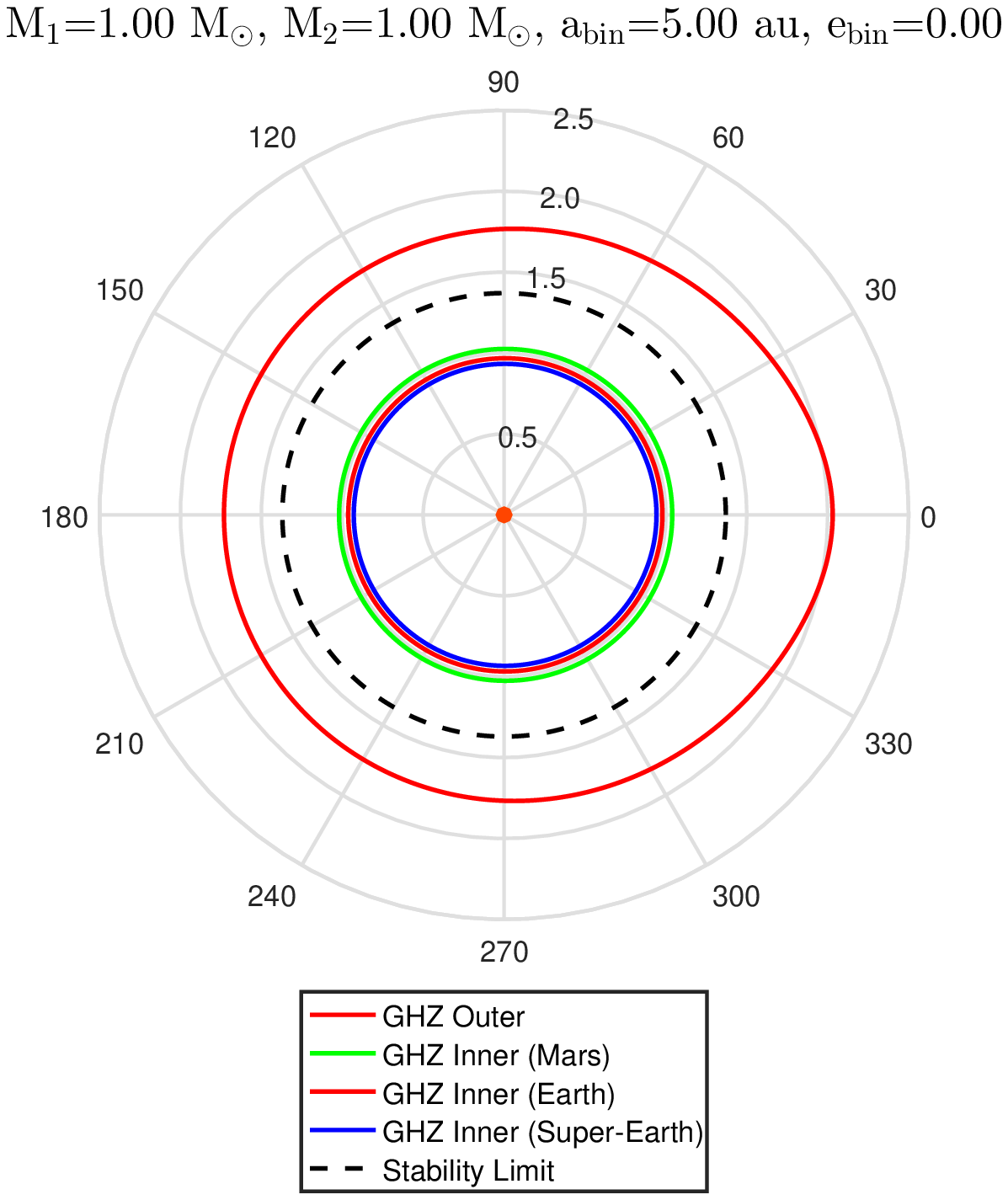,width=0.50\linewidth}
\epsfig{file=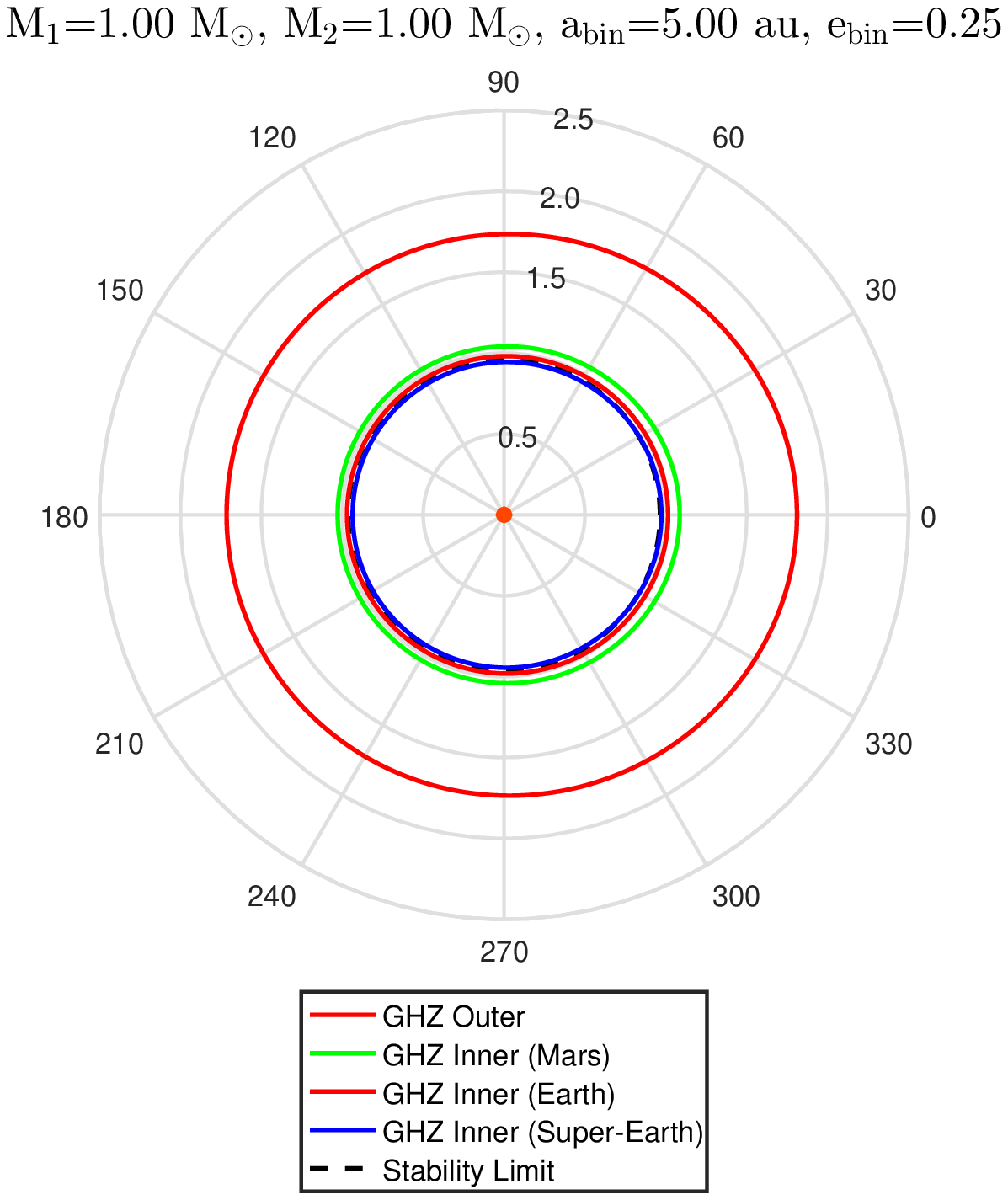,width=0.50\linewidth} \\
\epsfig{file=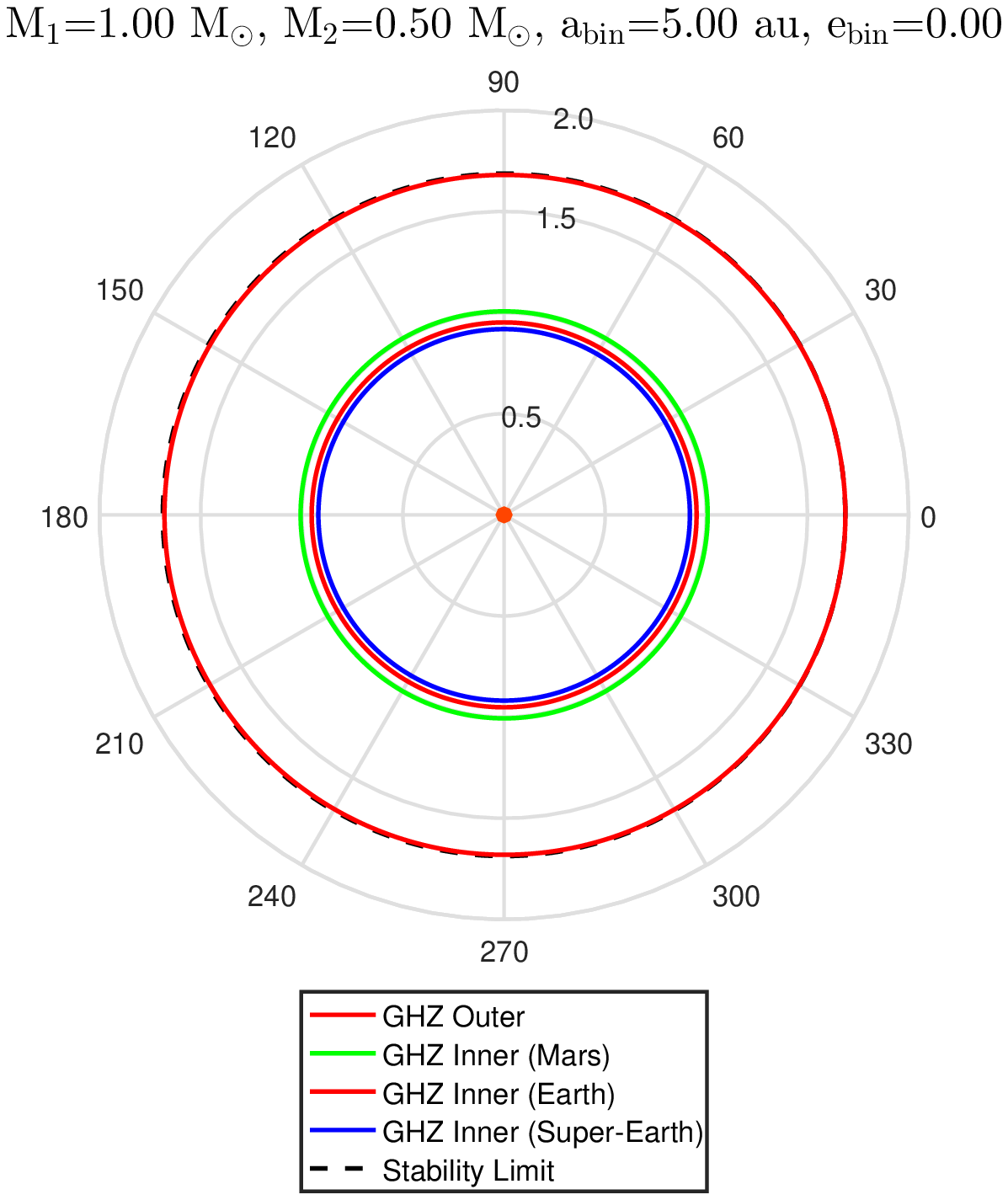,width=0.50\linewidth}
\epsfig{file=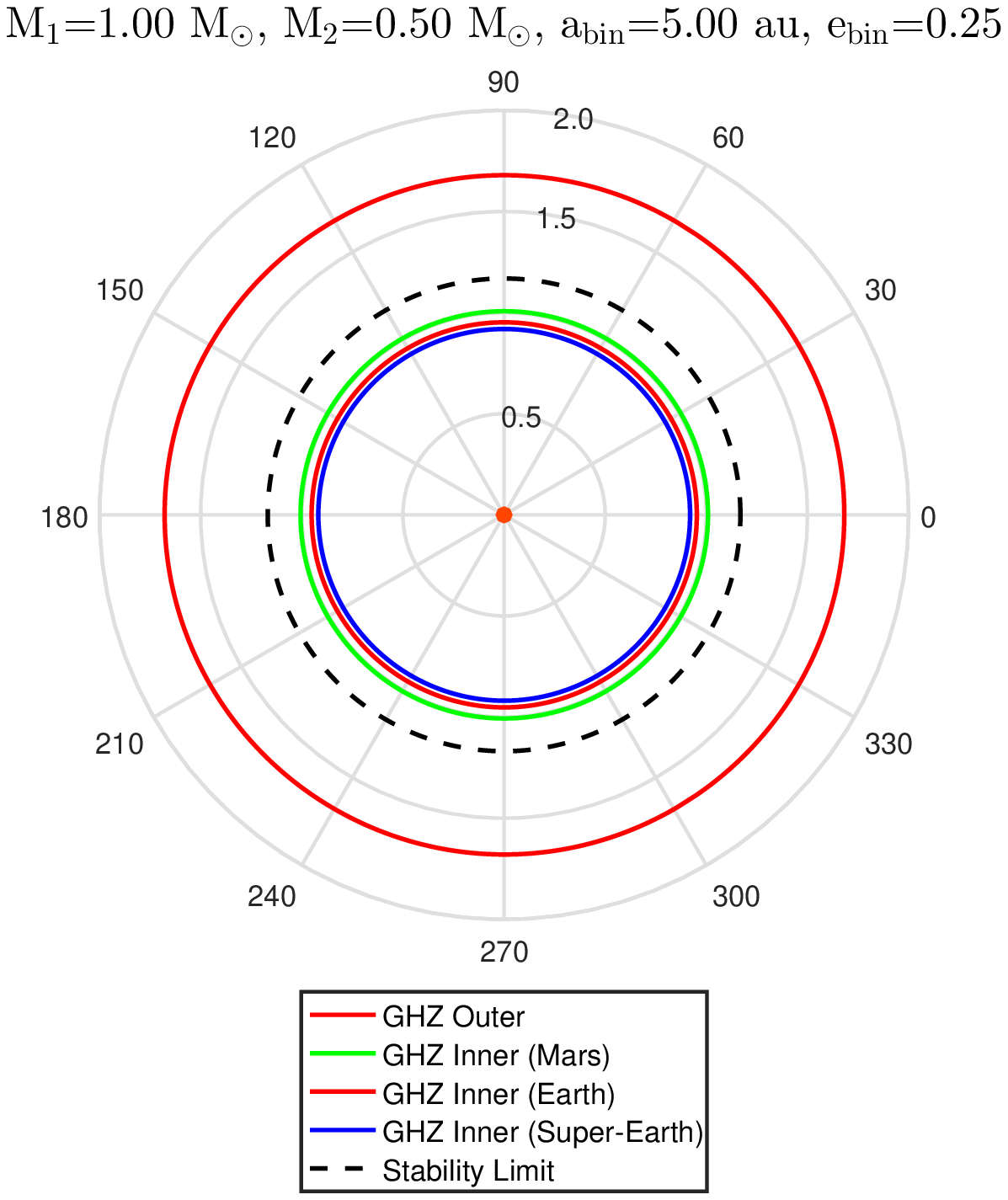,width=0.50\linewidth} \\
\end{tabular}
\caption{
Polar diagrams for {\it S}-type habitable zones for the GHZ, assuming Mars-type (0.1~$M_\oplus$), Earth-type,
and super-Earth-type (5.0~$M_\oplus$) planets, for two different combinations of stellar masses.
The stability limit (which is an outer limit) is given as well.
Here $a_{\rm bin}$ = 5~au is used.  Results are shown for $e_{\rm bin}$ = 0.00 and 0.25.
The secondary star is assumed to be to the right.
In the bottom figure (left), the lines for GHZ outer limit and planetary stability limit largely coincide.
}
\end{figure*}
\clearpage


\begin{figure*} 
\centering
\begin{tabular}{c}
\epsfig{file=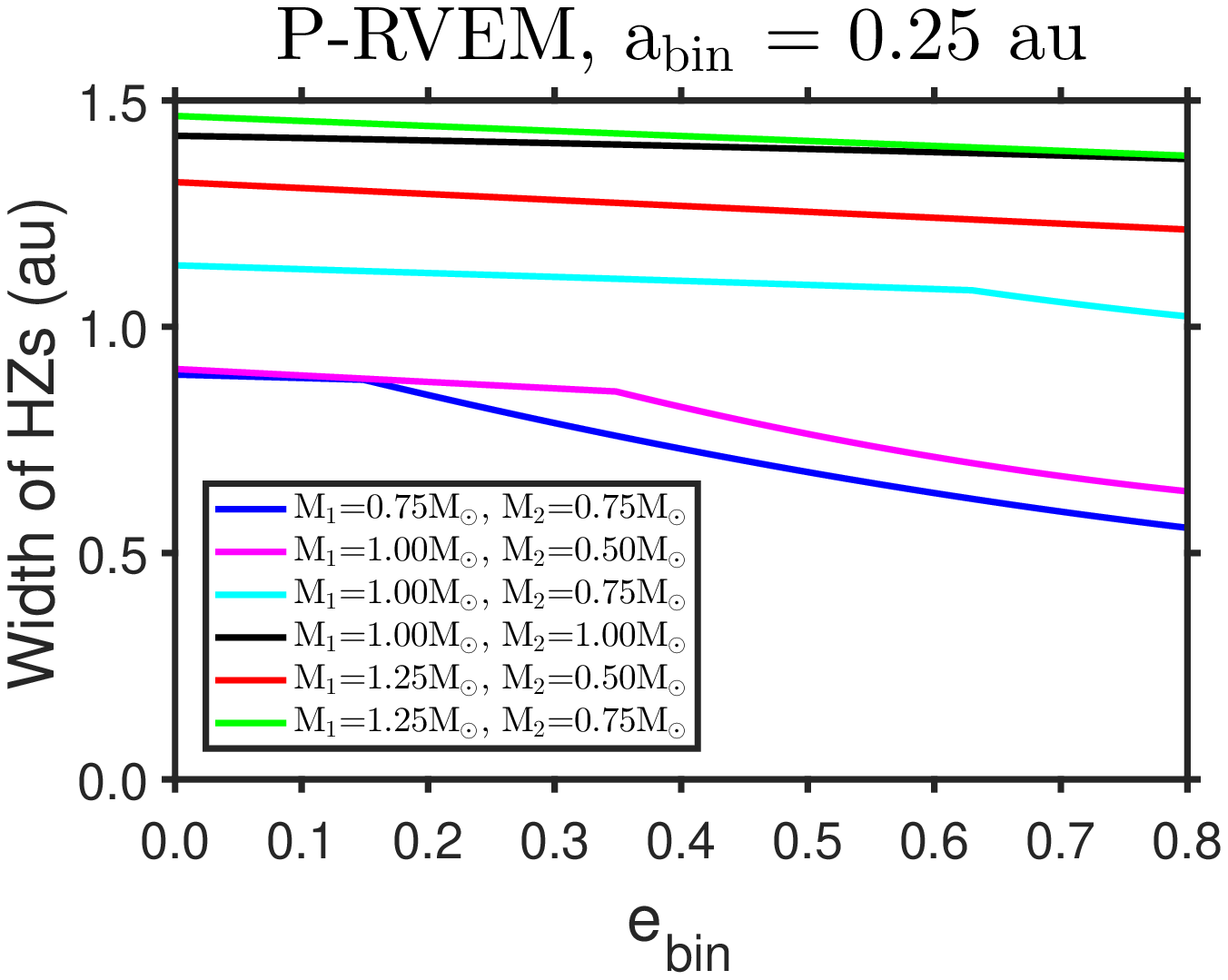,width=0.7\linewidth} \\
\epsfig{file=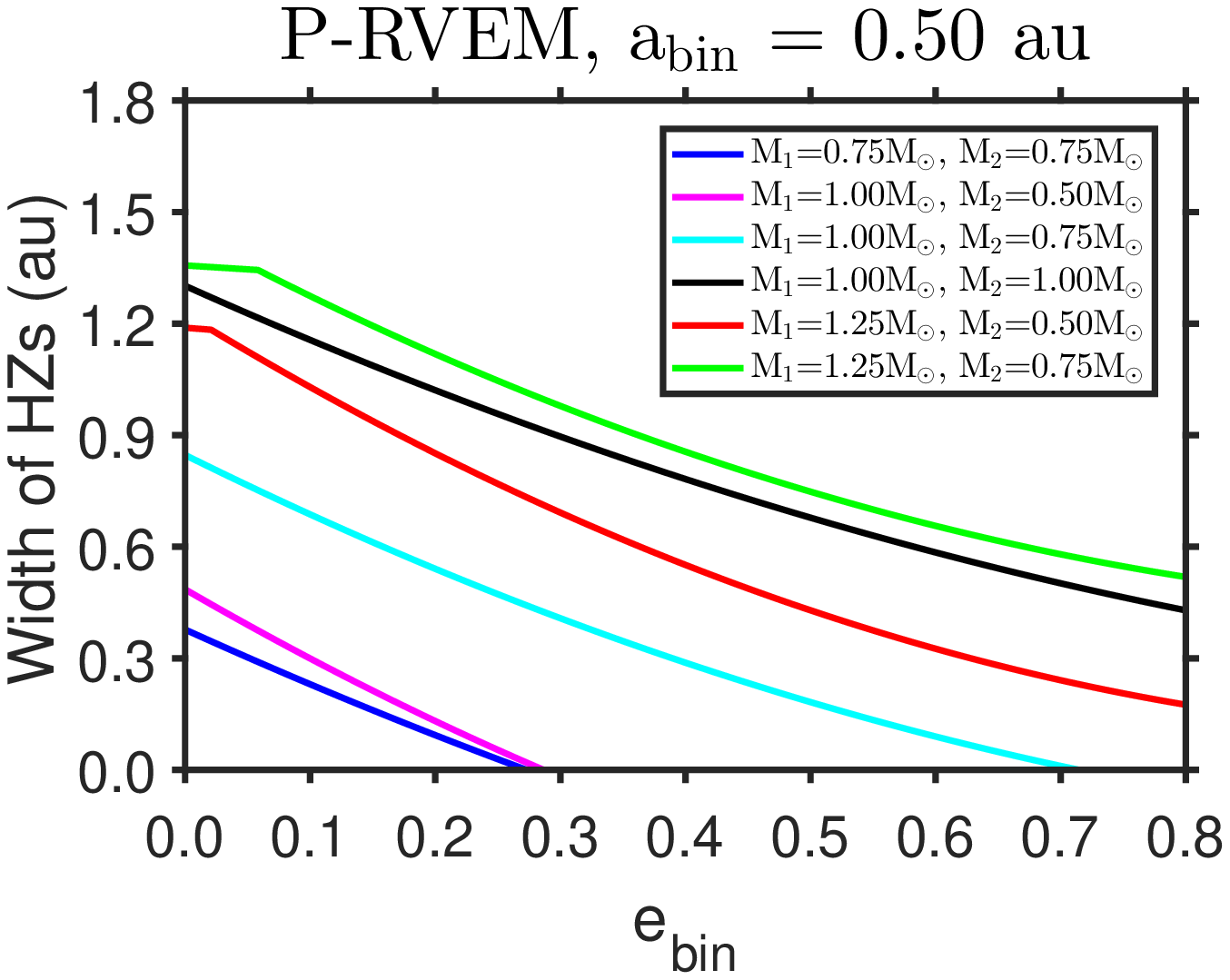,width=0.7\linewidth} \\
\end{tabular}
\caption{
Widths of {\it P}/{\it PT}-type habitable zones based on RVEM limits
for various binary systems.  Here we depict results for $a_{\rm bin}=0.25$~au
(top) and 0.50~au (bottom).
}
\end{figure*}
\clearpage


\begin{figure*}
\centering
\begin{tabular}{c}
\epsfig{file=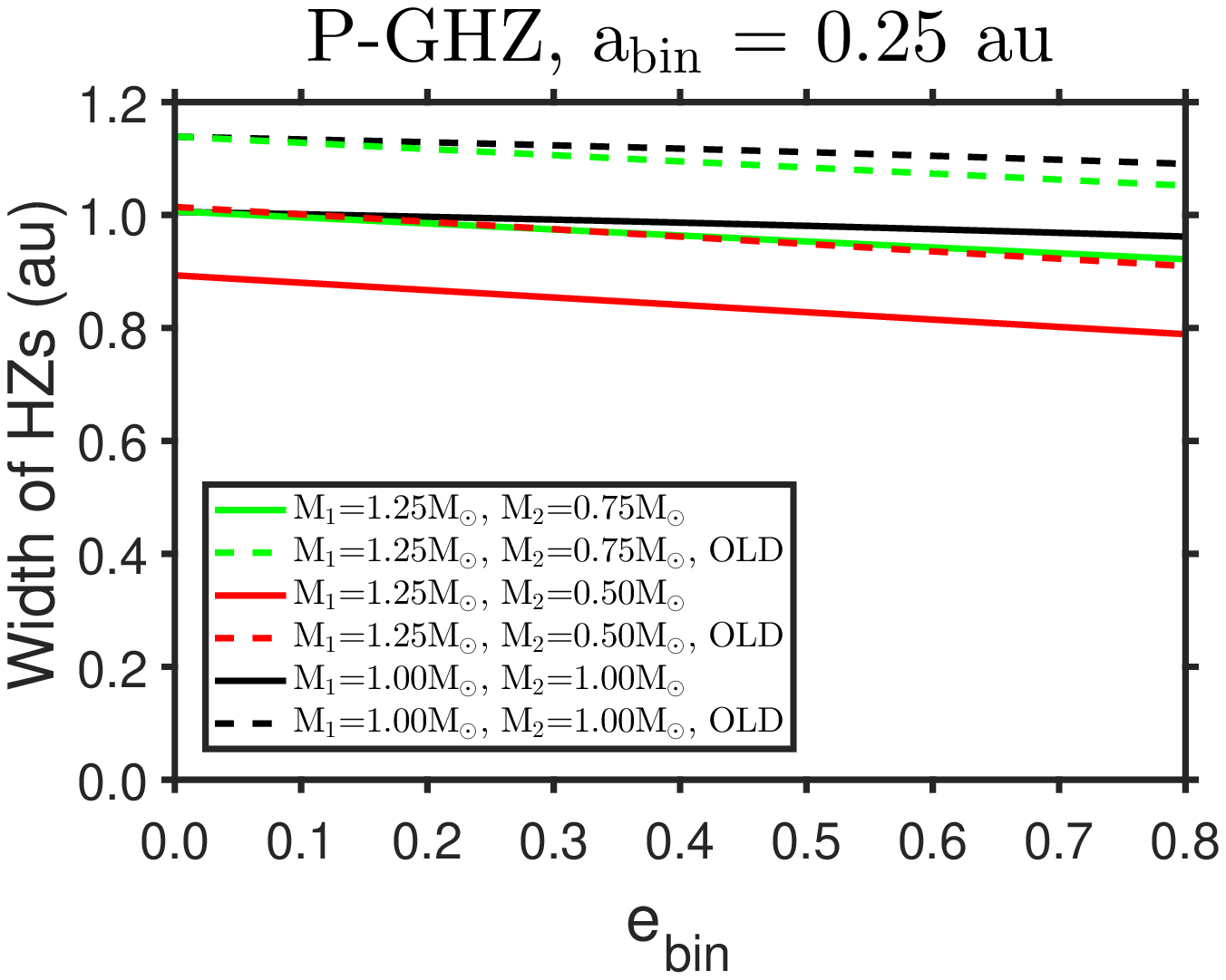,width=0.7\linewidth} \\
\epsfig{file=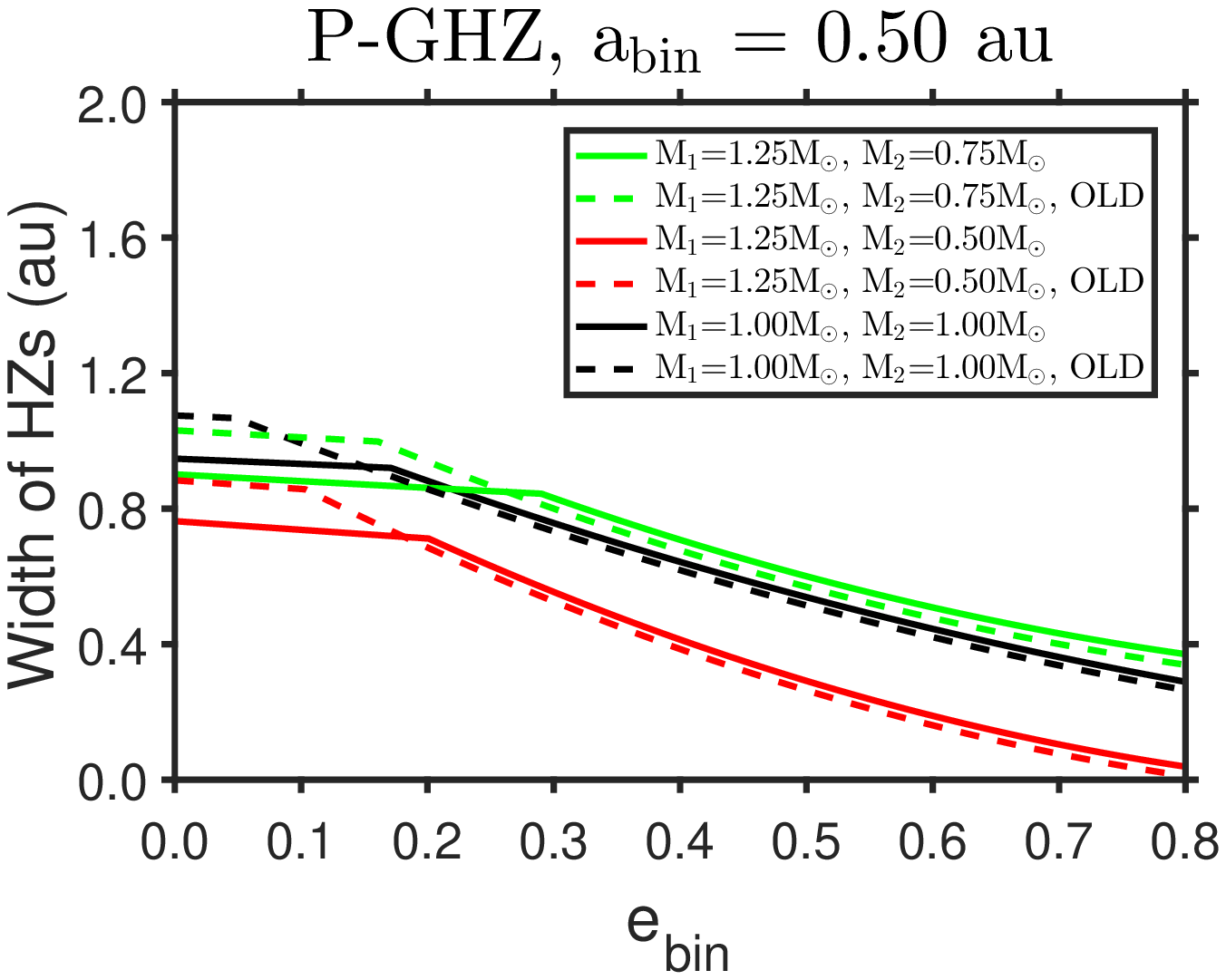,width=0.7\linewidth} \\
\end{tabular}
\caption{
Widths of {\it P}/{\it PT}-type habitable zones based on GHZ limits, based on Earth-type planets,
for various binary systems.  Here we depict results for $a_{\rm bin}=0.25$~au
(top) and 0.50~au (bottom).
The models tagged as ``OLD" are based on the effective stellar flux approximation by \cite{sel07}.
}
\end{figure*}
\clearpage


\begin{figure*} 
\centering
\begin{tabular}{c}
\epsfig{file=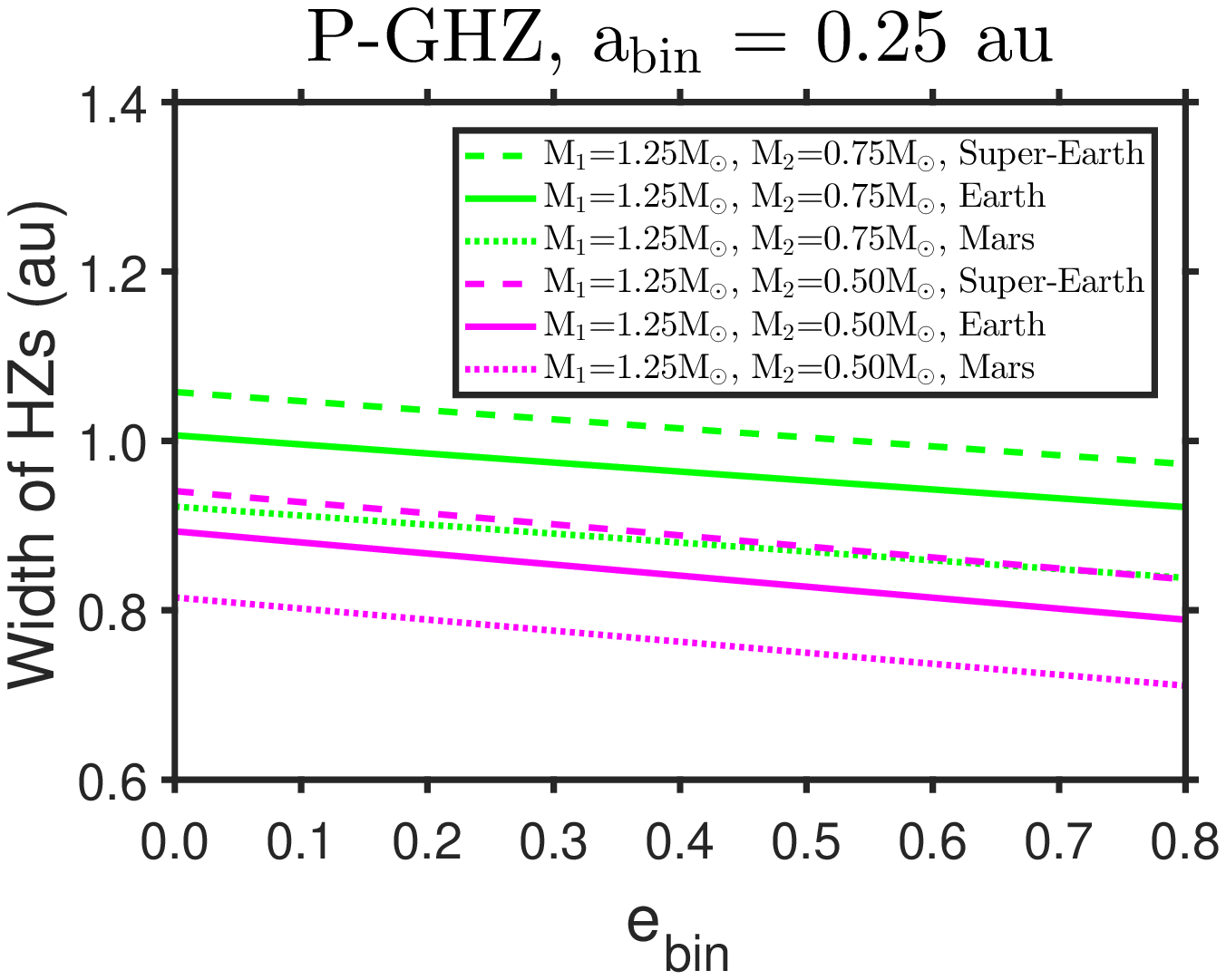,width=0.7\linewidth} \\
\epsfig{file=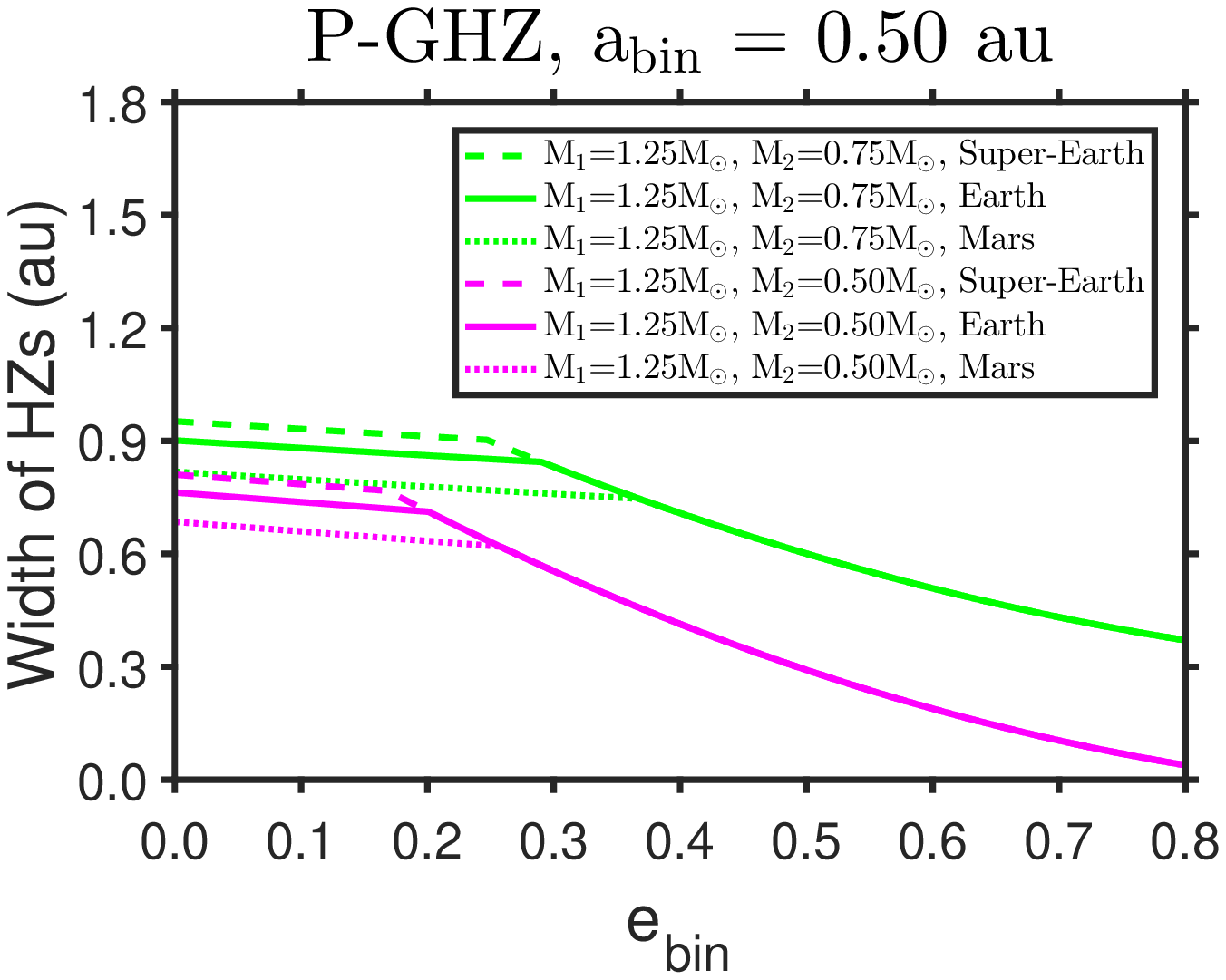,width=0.7\linewidth} \\
\end{tabular}
\caption{
Widths of {\it P}/{\it PT}-type habitable zones based on GHZ limits, based on
Mars-type (0.1~$M_\oplus$), Earth-type,
and super-Earth-type (5.0~$M_\oplus$) planets, for various binary systems. 
Here we depict results for $a_{\rm bin}=0.25$~au (top) and 0.50~au (bottom).
}
\end{figure*}
\clearpage


\begin{figure*} 
\centering
\begin{tabular}{c}
\epsfig{file=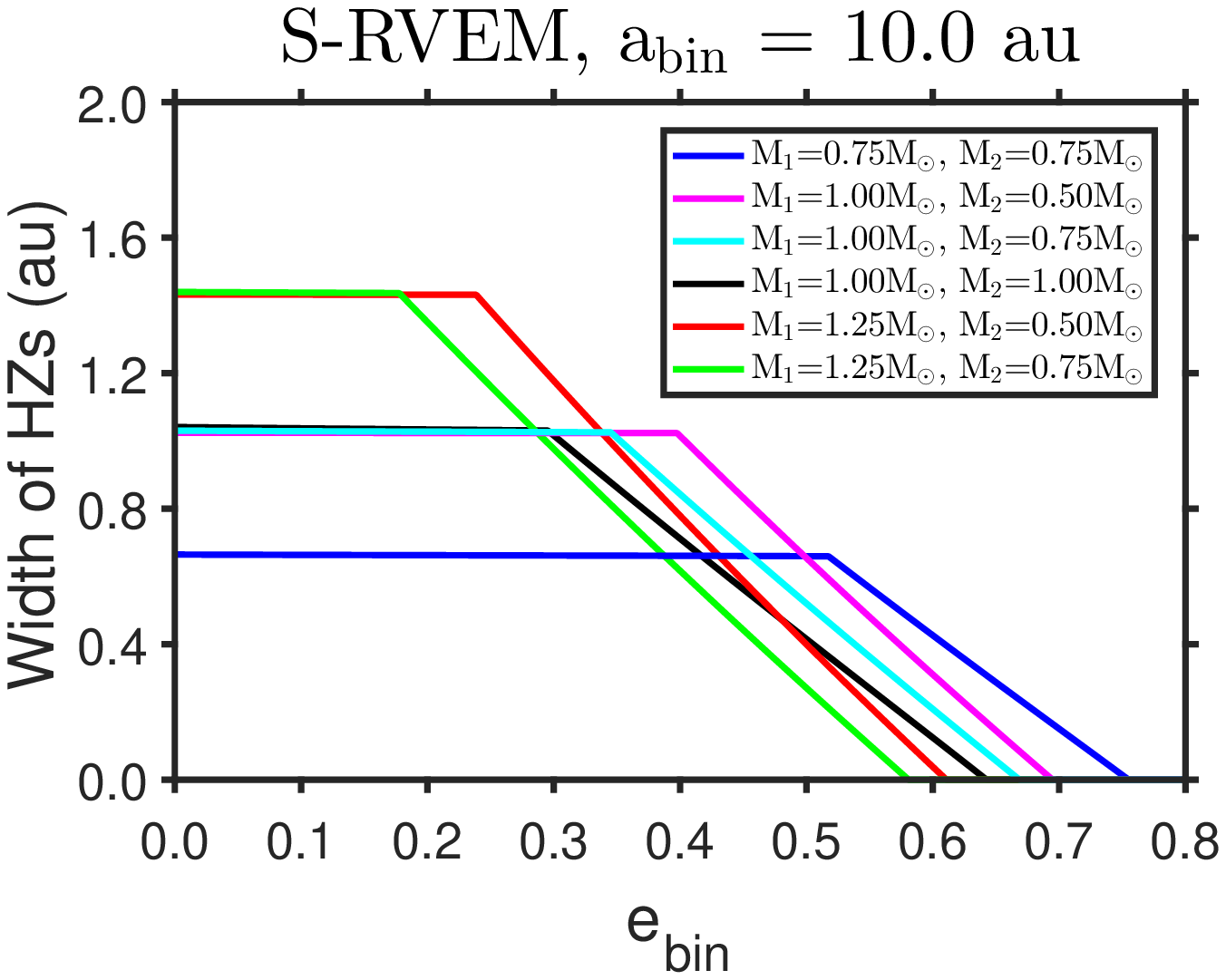,width=0.7\linewidth} \\
\epsfig{file=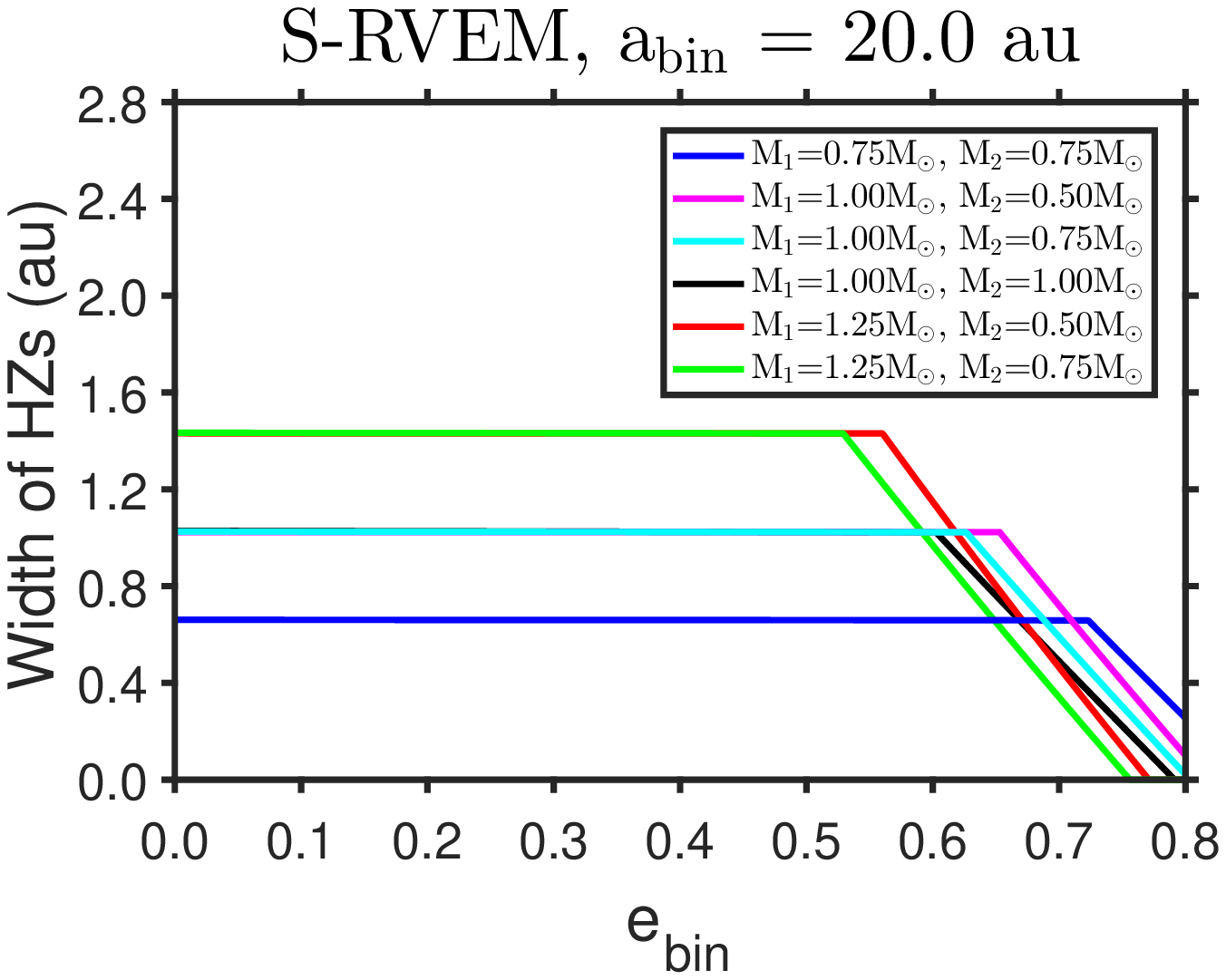,width=0.7\linewidth} \\
\end{tabular}
\caption{
Widths of {\it S}/{\it ST}-type habitable zones based on RVEM limits
for various binary systems.  Here we depict results for $a_{\rm bin}=10$~au
(top) and 20~au (bottom).
}
\end{figure*}
\clearpage


\begin{figure*} 
\centering
\begin{tabular}{c}
\epsfig{file=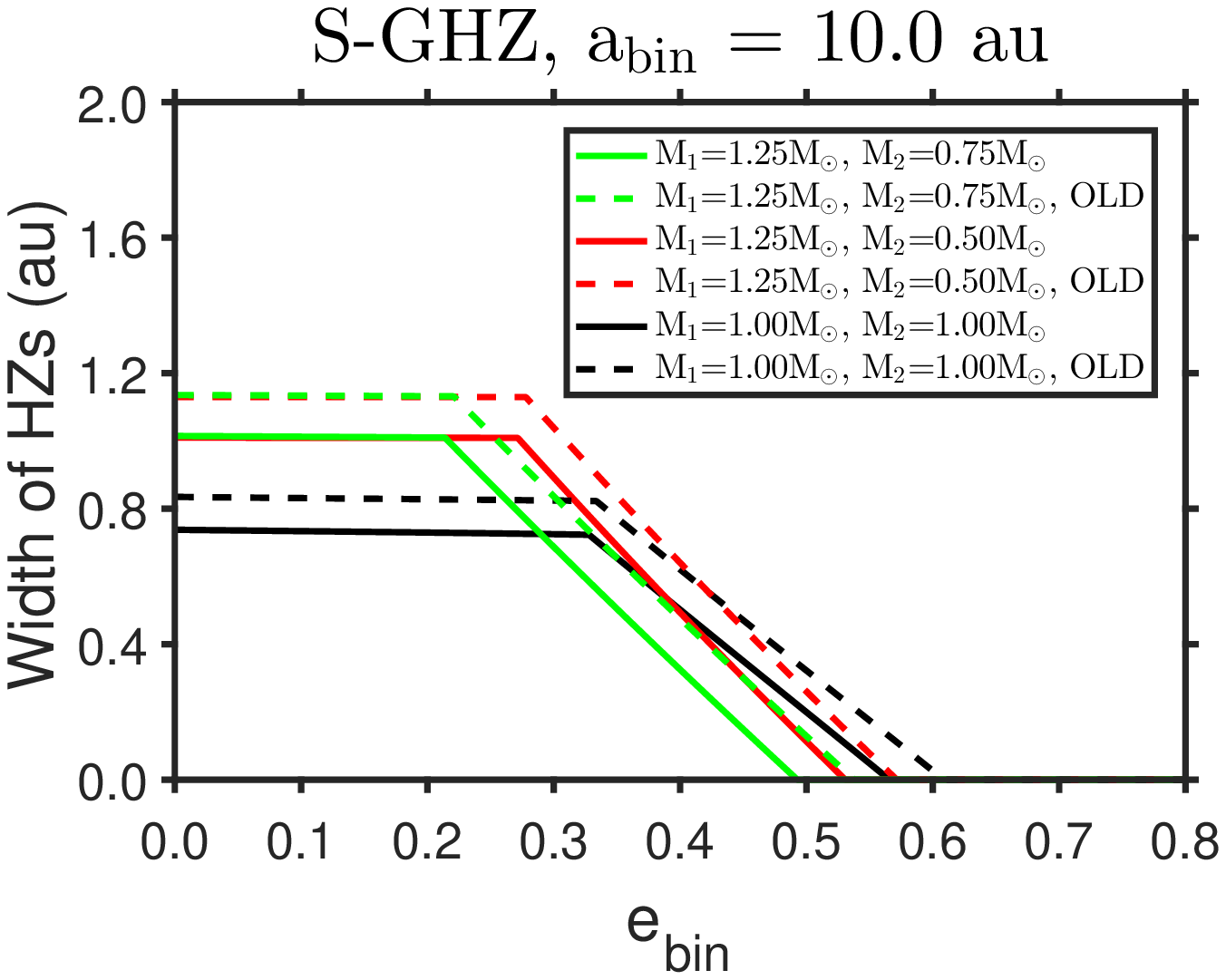,width=0.7\linewidth} \\
\epsfig{file=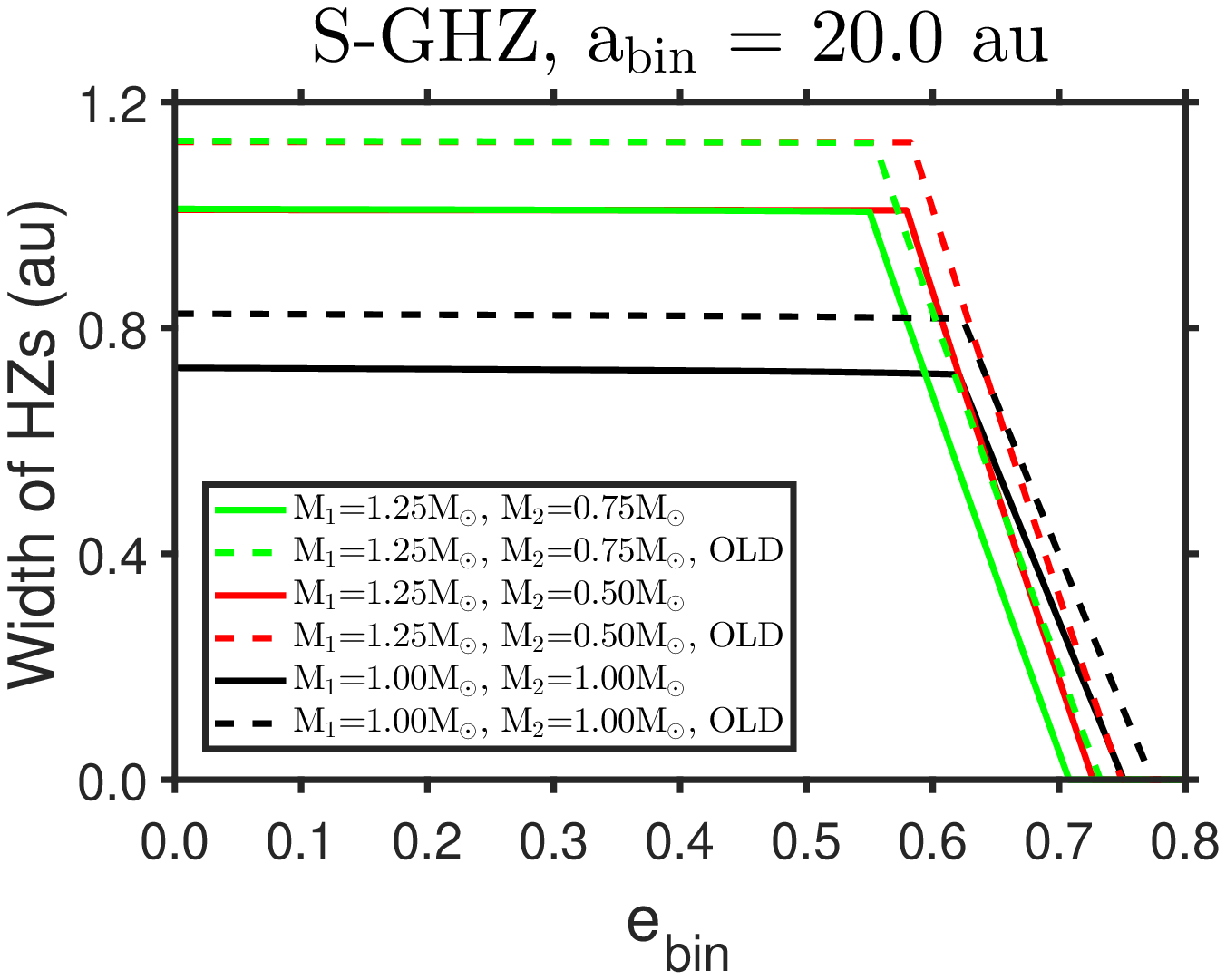,width=0.7\linewidth} \\
\end{tabular}
\caption{
Widths of {\it S}/{\it ST}-type habitable zones based on GHZ limits, based on Earth-type planets,
for various binary systems.  Here we depict results for $a_{\rm bin}=10$~au
(top) and 20~au (bottom).
The models tagged as ``OLD" are based on the effective stellar flux approximation by \cite{sel07}.
}
\end{figure*}
\clearpage


\begin{figure*} 
\centering
\begin{tabular}{c}
\epsfig{file=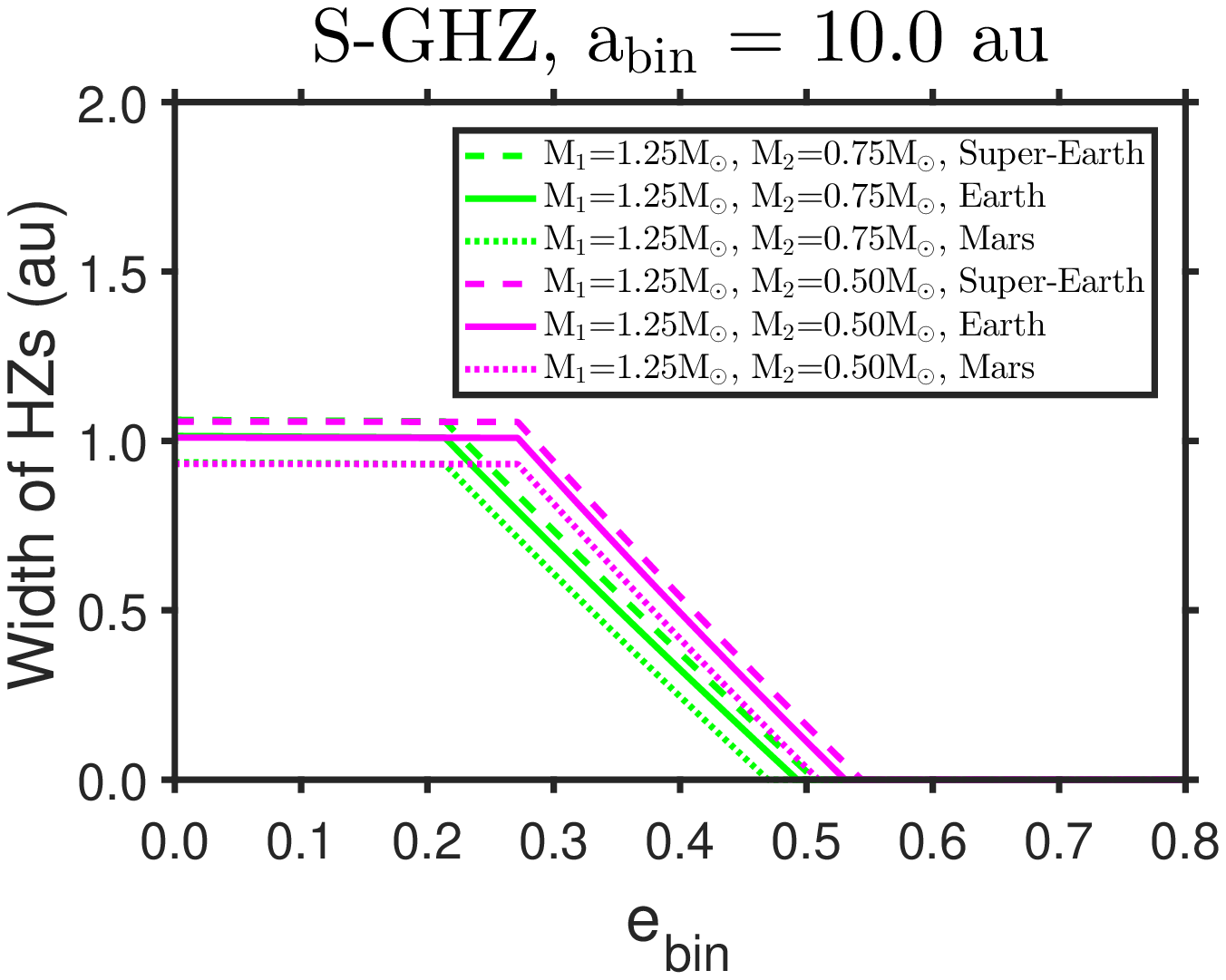,width=0.7\linewidth} \\
\epsfig{file=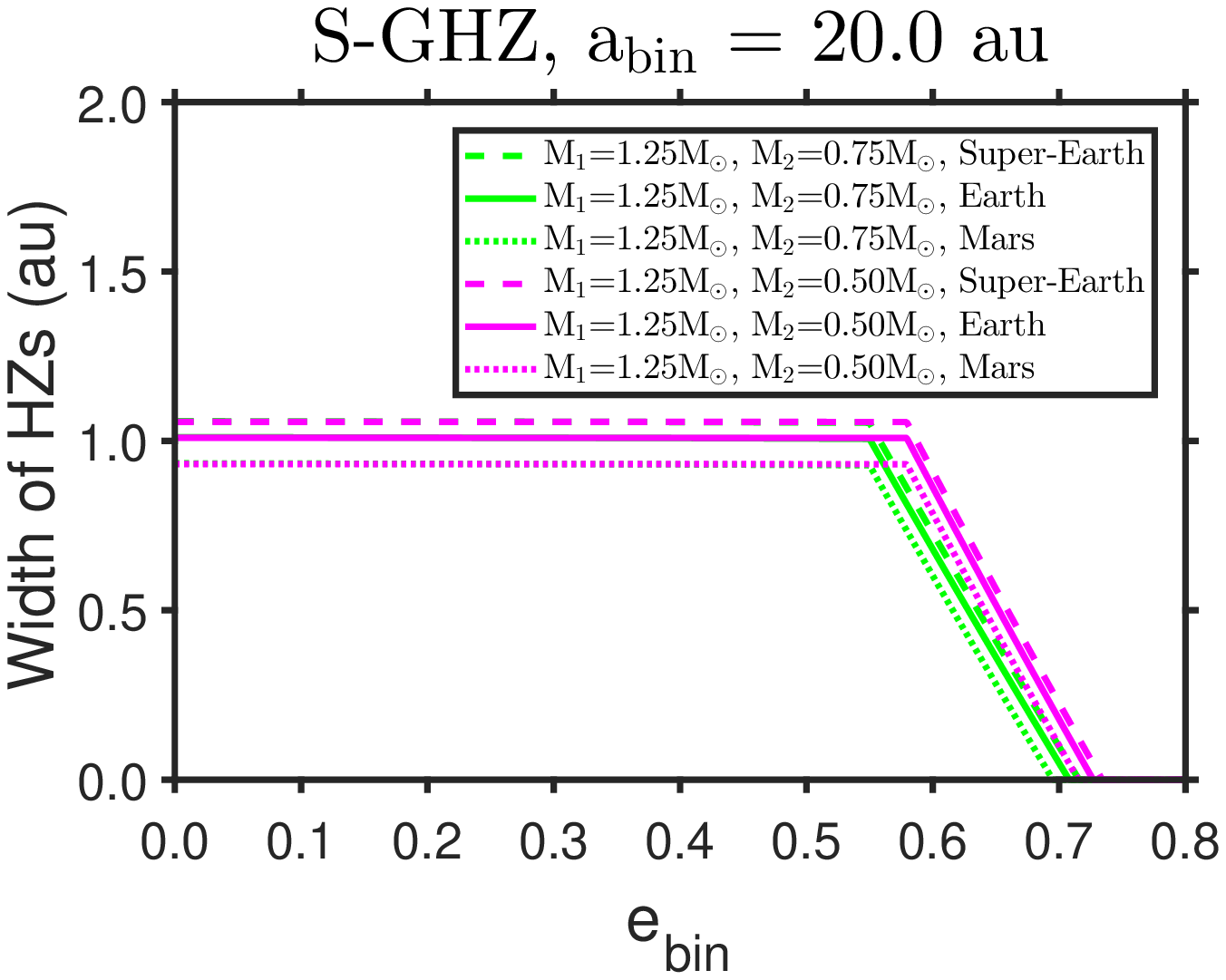,width=0.7\linewidth} \\
\end{tabular}
\caption{
Widths of {\it S}/{\it ST}-type habitable zones based on GHZ limits, based on
Mars-type (0.1~$M_\oplus$), Earth-type,
and super-Earth-type (5.0~$M_\oplus$) planets, for various binary systems.
Here we depict results for $a_{\rm bin}=10$~au (top) and 20~au (bottom).
}
\end{figure*}
\clearpage


\begin{deluxetable}{lccccc}
\tablecaption{Stellar Parameters}
\tablewidth{0pt}
\tablehead{
Sp. Type & $T_{\rm eff}$ & $R_\ast$    & $L_\ast$    & $M_\ast$ \\
       ... &  (K)        & ($R_\odot$) & ($L_\odot$) & ($M_\odot$)
}
\startdata
    F0       &  7178  &  1.62  &  6.24   &  1.60  \\ 
    F2       &  6909  &  1.48  &  4.47   &  1.52  \\ 
    F5       &  6528  &  1.40  &  3.19   &  1.40  \\ 
    F8       &  6160  &  1.20  &  1.86   &  1.19  \\ 
    G0       &  5943  &  1.12  &  1.40   &  1.05  \\ 
    G2       &  5811  &  1.08  &  1.19   &  0.99  \\ 
    G2~(Sun) &  5780  &  1.00  &  1.00   &  1.00  \\
    G5       &  5657  &  0.95  &  0.83   &  0.91  \\ 
    G8       &  5486  &  0.91  &  0.67   &  0.84  \\ 
    K0       &  5282  &  0.83  &  0.48   &  0.79  \\ 
    K2       &  5055  &  0.75  &  0.33   &  0.74  \\ 
    K4       &  4585  &  0.70  &  0.19   &  0.71  \\ 
    K5       &  4350  &  0.67  &  0.14   &  0.69  \\ 
    K6       &  4230  &  0.65  &  0.12   &  0.68  \\ 
    K8       &  4000  &  0.59  &  0.080  &  0.63  \\ 
    M0       &  3800  &  0.53  &  0.052  &  0.57  \\ 
    M1       &  3650  &  0.47  &  0.034  &  0.49  \\ 
    M2       &  3500  &  0.38  &  0.020  &  0.39  \\ 
\enddata
\end{deluxetable}
\clearpage


\begin{deluxetable}{lccccc}
\tablecaption{Habitability Limits for the Sun, $s_\ell$}
\tablewidth{0pt}
\tablehead{
Description & \multicolumn{2}{c}{Indices} & \multicolumn{3}{c}{Models}              \\
\noalign{\smallskip}
\hline
\noalign{\smallskip}
...         & $\ell$ & $k$ & \multicolumn{2}{c}{Kas93} &  Kop1314  \\
\noalign{\smallskip}
\hline
\noalign{\smallskip}
...         & ...    & ...  & 5700~K   & 5780~K   &  5780~K \\
...         & ...    & ...  & (au)     & (au)     &  (au)
}
\startdata
Recent Venus                            & 1 & 1 &  0.75   &  0.77  & 0.750  \\
Runaway greenhouse effect               & 2 & 1 &  0.84   &  0.86  & 0.950  \\
Runaway greenhouse effect               & 2 & 0 &  ...    &  ...   & 1.005  \\
Runaway greenhouse effect               & 2 & 2 &  ...    &  ...   & 0.917  \\
Moist greenhouse effect                 & 3 & 1 &  0.95   &  0.97  & 0.993  \\
Earth-equivalent position               & 0 & 1 &  0.993  &  \eq1  & \eq1   \\
First CO$_2$ condensation               & 4 & 1 &  1.37   &  1.40  & ...    \\
Maximum greenhouse effect, no clouds    & 5 & 1 &  1.67   &  1.71  & 1.676  \\
Early Mars                              & 6 & 1 &  1.77   &  1.81  & 1.768  \\
\enddata
\tablecomments{
Here 5700~K and 5780~K indicate the solar effective temperature adopted for the
respective model calculation.  The three-digit precision for some of the habitability
limits are mostly conveyed for tutorial reasons.  Moreover, the depictions for $\ell$,
indicating the inner / outer limits of the various types of HZs, do not always agree
with those of Paper~I and II.  Furthermore, the index {\it k} indicates Mars-type
(0.1~$M_\oplus$), Earth-type (1.0~$M_\oplus$), and super-Earth-type
(5.0~$M_\oplus$) planets, corresponding to 0, 1, and 2, respectively.
}
\end{deluxetable}
\clearpage



%
\begin{deluxetable}{lcccc}
\tablecaption{Types of Habitable Zones}
\tablewidth{0pt}
\tablehead{Description & Kas93 & Kop1314 & Paper~I \& II & This work
}
\startdata
Recent Venus                             &  RVEM  &  GHZ  &   ...      &  RVEM  \\
Runaway greenhouse effect                &  GHZ   &  CHZ  &   EHZ, GHZ &  GHZ   \\
Moist greenhouse effect                  &  CHZ   &  ...  &   CHZ      &  ...   \\
First CO$_2$ condensation                &  CHZ   &  ...  &   CHZ      &  ...   \\
Maximum greenhouse effect, no clouds     &  GHZ   &  CHZ  &   GHZ      &  GHZ   \\
Early Mars                               &  RVEM  &  GHZ  &   ...      &  RVEM  \\
Maximum greenhouse effect, 100\% clouds  &  ...   &  ...  &   EHZ      &  ...   \\
\enddata
\tablecomments{
In each column, the double appearance of terms as, e.g., GHZ indicates either an inner
or outer limit.  See main text for definitions and background information.
}
\end{deluxetable}
\clearpage


\begin{deluxetable}{lccccc}
\tablecaption{Main Target List}
\tablewidth{0pt}
\tablehead{
$M_\ast$    & Spectral Type & $T_{\rm eff}$ & $R_\ast$    & $L_\ast$    & HZ$(s_0)$ \\
($M_\odot$) & ...           & (K)           & ($R_\odot$) & ($L_\odot$) & (au) 
}
\startdata
 1.25  &  F8~V  &  6257  &  1.25  &  2.15   &  1.47  \\ 
 1.00  &  G2~V  &  5780  &  1.00  &  1.00   &  1.00  \\ %
 0.75  &  K2~V  &  5104  &  0.77  &  0.357  &  0.60  \\ 
 0.50  &  M1~V  &  3664  &  0.47  &  0.0359 &  0.19  \\ 
\enddata
\tablecomments{
HZ$(s_0)$ are calculated as the Earth-equivalent distance.
}
\end{deluxetable}
\clearpage


\begin{deluxetable}{lcccc}
\tablecaption{Habitability Classification for
              $M_1 = M_2 = 1.0~M_\odot$}
\tablewidth{0pt}
\tablehead{
$e_{\rm bin}$ & {\it P} & {\it PT} & {\it ST} & {\it S} \\
...         & (au)    & (au)     & (au)     & (au)
}
\startdata
\multicolumn{5}{c}{RVEM} \\
\noalign{\smallskip}
\hline
\noalign{\smallskip}
 0.00  &  0.474	&  1.028	&  2.928	&  6.620	\\
 0.25  &  0.369	&  0.802	&  4.121	&  9.285	\\
 0.50  &  0.314	&  0.681	&  6.669	&  15.100	\\
 0.75  &  0.284	&  0.610	&  15.707	&  36.066	\\
\noalign{\smallskip}
\hline
\noalign{\smallskip}
\multicolumn{5}{c}{GHZ} \\
\noalign{\smallskip}
\hline
\noalign{\smallskip}
 0.00  &  0.601	&  0.971	&  3.708	&  6.257	\\
 0.25  &  0.467	&  0.758	&  5.220	&  8.776	\\
 0.50  &  0.398	&  0.644	&  8.446	&  14.272	\\
 0.75  &  0.360	&  0.577	&  19.895	&  34.087	\\
\enddata
\tablecomments{
The values as obtained assume Earth-type planets.
The second column indicates the maximum $a_{\rm bin}$ for {\it P}-type HZs to exist at a given binary eccentricity
without the constraint of orbital stability.  Values between the second and the third column indicate {\it PT}-type HZs
to exist, i.e., the constraint of orbital stability applies.  Values between the third and fourth column indicate that
no HZs to exist.  Values between the fourth and the fifth column indicate {\it ST}-type HZs to exist, i.e.,
the constraint of orbital stability applies.  Values larger than those given in the fifth column indicate {\it S}-type HZs
to exist, i.e., without the constraint of orbital stability.
}
\end{deluxetable}
\clearpage


\begin{deluxetable}{lcccc}
\tablecaption{Habitability Classification for
              $M_1 = 1.25~M_\odot$, $M_2 = 0.75~M_\odot$}
\tablewidth{0pt}
\tablehead{
$e_{\rm bin}$ & {\it P} & {\it PT} & {\it ST} & {\it S} \\
...         & (au)    & (au)     & (au)     & (au)
}
\startdata
\multicolumn{5}{c}{RVEM} \\
\noalign{\smallskip}
\hline
\noalign{\smallskip}
 0.00  &  0.542	&  1.076	&  3.398	&  7.825	\\
 0.25  &  0.407	&  0.817	&  4.870	&  11.219	\\
 0.50  &  0.346	&  0.691	&  8.038	&  18.562	\\
 0.75  &  0.317	&  0.626	&  19.365	&  44.921	\\
\noalign{\smallskip}
\hline
\noalign{\smallskip}
\multicolumn{5}{c}{GHZ} \\
\noalign{\smallskip}
\hline
\noalign{\smallskip}
 0.00  	&  0.687	&  1.016	&  4.304	&  7.397	\\
 0.25  	&  0.516 	&  0.772	&  6.168	&  10.606	\\
 0.50  	&  0.438	&  0.653	&  10.180	&  17.547	\\
 0.75  	&  0.401	&  0.592	&  24.528	&  42.466	\\
\enddata
\tablecomments{See Table~5 for explanations.}
\end{deluxetable}
\clearpage


\begin{deluxetable}{ccccccccc}
\tablecaption{Critical Values of $e_{\rm bin}$ for $P/PT$-Type Habitability}
\tablewidth{0pt}
\tablehead{
Model                    &  \multicolumn{4}{c}{$a_{\rm bin} = 0.25$~au}  &
                            \multicolumn{4}{c}{$a_{\rm bin} = 0.50$~au}  \\
\noalign{\smallskip}
\hline
\noalign{\smallskip}
Habitability Limits      &
\multicolumn{2}{c}{RVEM} & \multicolumn{2}{c}{GHZ} & \multicolumn{2}{c}{RVEM} & \multicolumn{2}{c}{GHZ}
}
\startdata
$M_1 / M_2$ $(M_\odot / M_\odot)$ 	& {\it P} & {\it PT} & {\it P} & {\it PT} & {\it P} & {\it PT} & {\it P} & {\it PT} \\
\noalign{\smallskip}
\hline
\noalign{\smallskip}
  1.25 / 1.25  &   $\dag$  	&  $\dag$  	&  $\dag$  	&  $\dag$  	&   0.318 	&  $\dag$  	&  $\dag$  	&  $\dag$	\\
  1.25 / 1.00  &   $\dag$  	&  $\dag$  	&  $\dag$  	&  $\dag$  	&   0.150 	&  $\dag$  	&  0.474  	&  $\dag$	\\
  1.25 / 0.75  &   $\dag$  	&  $\dag$ 	&  $\dag$  	&  $\dag$  	&   0.058 	&  $\dag$  	&  0.290    	&  $\dag$	\\
  1.25 / 0.50  &   $\dag$  	&  $\dag$  	&  $\dag$  	&  $\dag$  	&   0.020 	&  $\dag$  	&  0.200   	&  $\dag$   	\\
  1.00 / 1.00  &   $\dag$  	&  $\dag$  	&  $\dag$  	&  $\dag$  	&   ...       	&  $\dag$  	&  0.171    	&  $\dag$  	\\
  1.00 / 0.75  &   0.630  	&  $\dag$  	&  $\dag$  	&  $\dag$  	&   ...    	&  0.714   	&  0.019   	&  0.569  	\\
  1.00 / 0.50  &   0.348  	&  $\dag$  	&  $\dag$  	&  $\dag$  	&   ...    	&  0.286     	&  ...    	&  0.221  	\\
  0.75 / 0.75  &   0.148  	&  $\dag$  	&  0.477  	&  $\dag$  	&   ...    	&  0.272    	&  ...     	&  0.202  	\\
  0.75 / 0.50  &   ...     	&  $\dag$ 	&  0.137   	&  0.763  	&   ...    	&  ...     	&  ...     	&  ... 		\\
  0.50 / 0.50  &   ...     	&  ...     	&  ...     	&  ...     	&   ...   	&  ...     	&  ...     	&  ...     	\\
\enddata
\tablecomments{($\dag$) means that the critical value of $e_{\rm bin}$
is larger than 0.80; it could not be determined owing to the limitations
of the work by HW99.  Ellipsis indicates that there is no solution, which means no HZ could be found.
The term ``critical value" means that it is the {\it maximal} possible value for $e_{\rm bin}$,
allowing $P$-type or $PT$-type habitability to exist.
The results are given for Earth-type planets.
}
\end{deluxetable}
\clearpage


\begin{deluxetable}{ccccccc}
\tablecaption{Critical Values of $e_{\rm bin}$ for $P$-Type Habitability, cont'd}
\tablewidth{0pt}
\tablehead{
Model                    &  \multicolumn{3}{c}{$a_{\rm bin} = 0.25$~au}  &
                            \multicolumn{3}{c}{$a_{\rm bin} = 0.50$~au}  \\
\noalign{\smallskip}
\hline
\noalign{\smallskip}
$M_1 / M_2$ $(M_\odot / M_\odot)$ 	&  Mars  &  Earth  &  Super-Earth &  Mars  &  Earth  &  Super-Earth
}
\startdata
  1.25 / 1.25  &  $\dag$ 	& $\dag$ 	& $\dag$ 	&  $\dag$ 	&  $\dag$ 	&  0.722	\\
  1.25 / 1.00  &  $\dag$ 	& $\dag$ 	& $\dag$ 	&  0.596 	&  0.474 	&  0.410	\\
  1.25 / 0.75  &  $\dag$ 	& $\dag$ 	& $\dag$ 	&  0.367 	&  0.290 	&  0.247	\\
  1.25 / 0.50  &  $\dag$ 	& $\dag$ 	& $\dag$ 	&  0.257  	&  0.200   	&  0.168  	\\
  1.00 / 1.00  &  $\dag$ 	& $\dag$ 	& $\dag$ 	&  0.235 	&  0.171	&  0.133	\\
  1.00 / 0.75  &  $\dag$ 	& $\dag$ 	& $\dag$ 	&  0.063   	&  0.019   	&  ...   	\\
  1.00 / 0.50  &  $\dag$ 	& $\dag$ 	& $\dag$ 	&  ...		&  ... 	 	&  ...    	\\
  0.75 / 0.75  &  0.596 	& 0.477	& 0.414 	&  ... 		&  ...   	&  ...    	\\
  0.75 / 0.50  &  0.194   	& 0.137     	& 0.105     	&  ...    	&  ...    	&  ...     	\\
  0.50 / 0.50  &  ...    	& ...    	& ...    	&  ...    	&  ...    	&  ...     	\\
\enddata
\tablecomments{Same as Table~7, but for Mars-type (0.1~$M_\oplus$), Earth-type,
and super-Earth-type (5.0~$M_\oplus$) planets.
}
\end{deluxetable}
\clearpage


\begin{deluxetable}{ccccccccc}
\tablecaption{Critical Values of $e_{\rm bin}$ for $S/ST$-Type Habitability}
\tablewidth{0pt}
\tablehead{
Model                    &  \multicolumn{4}{c}{$a_{\rm bin} = 10.0$~au}  &
                            \multicolumn{4}{c}{$a_{\rm bin} = 20.0$~au}  \\
\noalign{\smallskip}
\hline
\noalign{\smallskip}
Habitability Limits      &
\multicolumn{2}{c}{RVEM} & \multicolumn{2}{c}{GHZ} & \multicolumn{2}{c}{RVEM} & \multicolumn{2}{c}{GHZ}
}
\startdata
$M_1 / M_2$ $(M_\odot / M_\odot)$ 	& {\it S} & {\it ST} & {\it S} & {\it ST} & {\it S} & {\it ST} & {\it S} & {\it ST} \\
\noalign{\smallskip}
\hline
\noalign{\smallskip}
  1.25 / 1.25  &   0.057 	&  0.520   	&  0.101 &  0.415   	&   0.473 	&  0.727	&  0.497  	&  0.670   	\\
  1.25 / 1.00  &   0.118 	&  0.551   	&  0.157	&  0.454  	&   0.500  	&  0.741	&  0.522  	&  0.689   	\\
  1.25 / 0.75  &   0.177  	&  0.580   	&  0.213 	&  0.492  	&   0.529  	&  0.755	&  0.549  	&  0.707  	\\
  1.25 / 0.50  &   0.238 	&  0.610  	&  0.271	&  0.530  	&   0.560     & 0.771	&  0.579  	&  0.726   	\\
  1.00 / 1.00  &   0.295 	&  0.642   	&  0.327	&  0.565  	&   0.602 	&  0.791	&  0.620  	&  0.751   	\\
  1.00 / 0.75  &   0.345  	&  0.668  	&  0.374 	&  0.599  	&   0.626 	&  $\dag$	&  0.642  	&  0.766   	\\
  1.00 / 0.50  &   0.397  	&  0.694   	&  0.424 	&  0.632  	&   0.652 	&  $\dag$	&  0.667  	&  0.783   	\\
  0.75 / 0.75  &   0.517 	&  0.754   	&  0.539 	&  0.705  	&   0.723  	&  $\dag$	&  0.734  	&  $\dag$   	\\
  0.75 / 0.50  &   0.558 	&  0.776   	&  0.578	&  0.733  	&   0.743  	&  $\dag$	&  0.754  	&  $\dag$   	\\
  0.50 / 0.50  &   0.785	&  $\dag$ 	&  0.794 	&  $\dag$   	&  $\dag$	&  $\dag$	&  $\dag$	&  $\dag$ 	\\
\enddata
\tablecomments{($\dag$) means that the critical value of $e_{\rm bin}$
is larger than 0.80; it could not be determined owing to the limitations
of the work by HW99.  Ellipsis indicates that there is no solution, which means no HZ could be found.
The term ``critical value" means that it is the {\it maximal} possible value for $e_{\rm bin}$,
allowing $S$-type or $ST$-type habitability to exist.
The results are given for Earth-type planets.
}
\end{deluxetable}
\clearpage


\begin{deluxetable}{ccccccc}
\tablecaption{Critical Values of $e_{\rm bin}$ for $ST$-Type Habitability, cont'd}
\tablewidth{0pt}
\tablehead{
Model                    &  \multicolumn{3}{c}{$a_{\rm bin} = 10.0$~au}  &
                            \multicolumn{3}{c}{$a_{\rm bin} = 20.0$~au}  \\
\noalign{\smallskip}
\hline
\noalign{\smallskip}
$M_1 / M_2$ $(M_\odot / M_\odot)$            &  Mars  &  Earth  &  Super-Earth &  Mars  &  Earth  &  Super-Earth
}
\startdata
  1.25 / 1.25  &  0.387   	&  0.415  	&  0.432   	&  0.655  &  0.670 	&  0.680 	\\
  1.25 / 1.00  &  0.429  	&  0.454   	&  0.470  	&  0.675	 &  0.689 	&  0.697 	\\
  1.25 / 0.75  &  0.469  	&  0.492  	&  0.507  	&  0.695	 &  0.707 	&  0.715 	\\
  1.25 / 0.50  &  0.509   	&  0.530  	&  0.543  	&  0.715	 &  0.726 	&  0.734 	\\
  1.00 / 1.00  &  0.545   	&  0.565  	&  0.578  	&  0.740	 &  0.751 	&  0.757 	\\
  1.00 / 0.75  &  0.580  	&  0.599  	&  0.610  	&  0.756	 &  0.766 	&  0.772 	\\
  1.00 / 0.50  &  0.616   	&  0.632  	&  0.642  	&  0.774	 &  0.783 	&  0.788 	\\
  0.75 / 0.75  &  0.691   	&  0.705  	&  0.713  	&  $\dag$ &  $\dag$ 	&  $\dag$ 	\\
  0.75 / 0.50  &  0.721   	&  0.733  	&  0.740  	&  $\dag$ &  $\dag$ 	&  $\dag$ 	\\
  0.50 / 0.50  &  $\dag$ 	&  $\dag$  	&  $\dag$ 	&  $\dag$ &  $\dag$ 	&  $\dag$ 	\\
\enddata
\tablecomments{Same as Table~9, but for Mars-type (0.1~$M_\oplus$), Earth-type,
and super-Earth-type (5.0~$M_\oplus$) planets.
}
\end{deluxetable}
\clearpage


\begin{deluxetable}{cccccc}
\tablecaption{Stellar Parameter Comparison}
\tablewidth{0pt}
\tablehead{
Model & Sp.~Type & $M$ & $T_{\rm eff}$ & $R_\ast$ & $L_\ast$ \\
\noalign{\smallskip}
\hline
\noalign{\smallskip}
... & ...       & $(M_\odot)$ & (K) & $(R_\odot)$ & $(L_\odot)$
}
\startdata
\cite{man13}  & $\sim$M5~V   & 0.2  &  3292  & 0.233 & $5.727\times10^{-3}$ \\
\cite{bar15}  & $\sim$M5~V   & 0.2  &  3262 & 0.218 & $4.786\times10^{-3}$ \\
\enddata
\end{deluxetable}
\clearpage


\begin{deluxetable}{ccccccc}
\tablecaption{Equal Mass Systems: $M_1 = M_2 = 0.20~M_\odot$}
\tablewidth{0pt}
\tablehead{
Habitable Zone 	& $a_{\rm bin}$                 & $e_{\rm bin}$       	& \multicolumn{4}{c}{Stellar Model}\\
\hline
			&                               &				& \multicolumn{2}{c}{\cite{man13}} 	& \multicolumn{2}{c}{\cite{bar15}}\\
\hline
 			&                               & 				& Inner Limit	& Outer Limit 	& Inner Limit	& Outer Limit \\
 			&  (au)                         &         		& (au)		& (au)		& (au)		& (au)
}
\startdata
{\it P}-type GHZ 		& 0.01			& 0.00				& 0.1115 	& 0.2180		& 0.1020 	& 0.1996 \\
{\it P}-type GHZ 		& 0.01 			& 0.40				& 0.1118 	& 0.2180		& 0.1024 	& 0.1996 \\
{\it P}-type GHZ 		& 0.05 			& 0.00				& 0.1188 	& 0.2166		& 0.1099 	& 0.1981 \\
{\it P}-type GHZ 		& 0.05 			& 0.40				& 0.1252 	& 0.2153		& 0.1166 	& 0.1966 \\
\hline
{\it S}-type GHZ 		& 2.50			& 0.00				& 0.0786	& 0.1545 		& 0.0719 	& 0.1414 \\
{\it S}-type GHZ 		& 2.50			& 0.40				& 0.0787	& 0.1543 		& 0.0720 	& 0.1413 \\
{\it S}-type GHZ 		& 5.00 			& 0.00				& 0.0786	& 0.1543 		& 0.0719 	& 0.1413 \\
{\it S}-type GHZ 		& 5.00 			& 0.40				& 0.0786	& 0.1542 		& 0.0719 	& 0.1412 \\
\enddata
\tablecomments{
Here the RHZ limits are listed for different systems. The {\it P}-type limits are given from the mass center of the binary star system,
whereas the {\it S}-type limits are given from the center of the primary.  Results are given for the GHZ based on the work by
\cite{kop13,kop14}, assuming a planet of one Earth mass.  The high precision for the data has been given for mostly tutorial reasons.
\cite{bar15} present M-star models of different ages.  Here we assume 5~Gyr, a data that is highly inconsequential.
}
\end{deluxetable}
\clearpage


\begin{deluxetable}{ccccccc}
\tablecaption{Non-Equal Mass Systems: $M_1 = 1.00~M_\odot$, $M_2 = 0.20~M_\odot$}
\tablewidth{0pt}
\tablehead{
Habitable Zone 	& $a_{\rm bin}$                 & $e_{\rm bin}$       	& \multicolumn{4}{c}{Stellar Model}\\
\hline
			&                               &				& \multicolumn{2}{c}{\cite{man13}} 	& \multicolumn{2}{c}{\cite{bar15}}\\
\hline
 			&                               & 				& Inner Limit	& Outer Limit 	& Inner Limit	& Outer Limit \\
 			&  (au)                         & 		           	& (au)		& (au)	      & (au)		& (au)
}
\startdata
{\it P}-type GHZ 		& 0.01			& 0.00				& 0.9553 	& 1.6815 		& 0.9548 	& 1.6803 \\
{\it P}-type GHZ 		& 0.01 			& 0.40				& 0.9559 	& 1.6809 		& 0.9554 	& 1.6797 \\
{\it P}-type GHZ 		& 0.05  			& 0.00				& 0.9617 	& 1.6752 		& 0.9612 	& 1.6740 \\
{\it P}-type GHZ 		& 0.05  			& 0.40				& 0.9649 	& 1.6720 		& 0.9645 	& 1.6708 \\
\hline
{\it S}-type GHZ 		& 5.00  			& 0.00				& 0.9506 	& 1.6765 		& 0.9506 	& 1.6764 \\
{\it S}-type GHZ 		& 5.00  			& 0.40				& 0.9511 	& 1.6763 		& 0.9510 	& 1.6762 \\
{\it S}-type GHZ 		& 6.00 			& 0.00				& 0.9506 	& 1.6763 		& 0.9505 	& 1.6763 \\
{\it S}-type GHZ 		& 6.00 			& 0.40				& 0.9509 	& 1.6762 		& 0.9508 	& 1.6762 \\
\enddata
\tablecomments{
See Table~12 for comments.
}
\end{deluxetable}

\end{CJK*}
\end{document}